\documentclass[11pt]{article}

\usepackage[english]{babel}
\usepackage{standalone}
\usepackage[T1]{fontenc}
\usepackage[singlespacing]{setspace}
\usepackage[a4paper,top=2.5cm,bottom=2cm,left=2.4cm,right=2.4cm]{geometry}

\usepackage{amsmath, amssymb, amsthm, amsbsy, amsfonts, natbib, wrapfig,graphicx, comment, enumitem, algorithm, algcompatible,indentfirst} 
\usepackage[colorinlistoftodos]{todonotes}
\usepackage[colorlinks=true, allcolors=blue]{hyperref}
\usepackage[noend]{algpseudocode}
\usepackage{caption}
\usepackage{colortbl, titlesec}
\usepackage[font=small,skip=0pt]{caption}

\theoremstyle{plain}

\newtheorem{cor}{Corollary}
\newtheorem{pro}{Proposition}
\newtheorem{rem}{Remark}

\newcommand{\x}{\mathbf{x}}
\newcommand{\X}{\mathbf{X}}

\newcommand{\e}{\mathbf{e}}

\newcommand{\E}{\mbox{E}}
\newcommand{\V}{\mbox{V}}
\newcommand{\A}{\mathbf{A}}

\newcommand{\bs}{\boldsymbol}

\newcommand{\1}{\mathbf{1}}

\newcommand{\diag}{\mbox{diag}}

\allowdisplaybreaks
\titlespacing*{\section}
{0pt}{6pt plus 3pt minus 3pt}{1pt plus 1pt}
\titlespacing*{\subsection}
{0pt}{6pt plus 3pt minus 3pt}{1pt plus 1pt}

\title{\Large AdaPtive Noisy Data Augmentation (PANDA) for Simultaneous Construction of Multiple Graph Models}
\author{\normalsize Yinan Li$^{1}$, Xiao Liu$^{2}$, and Fang Liu$^1$\footnote{Corresponding author email: fang.liu.131@nd.edu}\\
\small$^1$ Department of Applied and Computational Mathematics and Statistics\\
\small$^2$ Department of Psychology\\
\small University of Notre Dame, Notre Dame, IN 46556, U.S.A. }
\date{}

\begin{document}
\maketitle
\begin{abstract}	
We extend the data augmentation technique PANDA by Li et al.\ (2018) that regularizes single graph estimation to jointly learning multiple graphical models with various node types in a unified framework. We design two types of noise to augment the observed data: the first type regularizes the estimation of each graph while the second type promotes either the structural similarity, referred as the \emph{joint group lasso} regularization, or the numerical similarity, referred as the \emph{joint fused ridge} regularization, among the edges in the same position across graphs. The computation in PANDA is straightforward and only involves obtaining maximum likelihood estimator in generalized linear models in an iterative manner.  The simulation studies demonstrate PANDA is non-inferior to existing joint estimation approaches for Gaussian graphical models, and significantly improves over the na\"{i}ve differencing approach for non-Gaussian graphical models. We apply PANDA to a real-life lung cancer microarray data to simultaneously construct four protein networks.

\vspace{6pt}

\noindent \textbf{keywords}: adjacency matrix, joint group lasso (JGL),  joint fused ridge (JFR), sparse group lasso (SGL),  sparse fused ridge (SFR),  sparsity and similarity
\end{abstract}

\pagebreak

\section{Introduction}\label{sec:intro}
Undirected graphical models (UGM) are often used to capture the bilateral conditional dependency structure among a set of variables called nodes. The structure can be mathematically expressed by a $p\times p$ symmetric adjacency matrix $\A$. If  entry $\A[i,j]$ is 0, then nodes $i$ and $j$  are conditionally independent; otherwise, there is conditional dependency and an edge is drawn  between the two nodes in the graph. In addition to the extensive research and work on  single graph construction 
there are also increasing practical needs and  research interests in comparing multiple graphs, the data of which are collected under different conditions. For example, medical researchers  collect and compare gene expression data from normal tissues and tumor tissues to learn how the connection pattern and network structure among the genes (nodes) change. Real-life graphs are often expected to take on sparse structures, especially for large-scale networks. It is also expected that, in the case of multiple graphs, the change in condition would only affect a small subset of edges while leaving the majority of connection patterns intact across the graphs. To meet the real-life expectations,  sparsity-promoting regularization is often applied when constructing both single and multiple graphs.

In the single graph setting, neighborhood selection (NS) is a popular and simple method to recover the structure of a UGM that guarantees the structure consistency \citep{yang2012,UGMEXP2015} when the underlying  conditional distribution of one node given all others can be modeled by an exponential family. NS transforms the original graph construction problem into building $p$ generalized linear models (GLM), where sparsity regularization, such as $l_1$,  can be applied directly. Methodology and theory have been developed in linear regression for Gaussian graphical models (GGM) \citep{Yuan2010}; logistic regression for Ising models  \citep{ising2010}, Bernoulli nodes \citep{binary2009}, and categorical nodes \citep{multinomial2011, Kuang2017}; and  Poisson regression for count nodes \citep{poisson2012}; as well as mixed graph models (MGM) when there are different node types in a UGM \citep{yang2012,mixed2013,yang2014}.  \citet{panda1} also provide the a data augmentation approach, named PANDA, for constructing a single UGM that offers a wide spectrum of regularization effects with properly designed augmented noise terms.

For multiple graph construction, a na\"{i}ve approach is estimating each graph separately and then comparing the structures in a post-hoc manner for differences. A better approach is to  jointly estimate multiple graphs simultaneously, acknowledging and leveraging the expectation that the structures would largely remain consistent across the multiple graphs while the discrepancy is in the minority. For example, \citet{Danaher2014} propose the fused graphical lasso and group graphical lasso 
to estimate the precision matrices of multiple GGMs simultaneously. \citet{Yang2015} consider the sequential version of the fused graphical lasso for training time-evolving GGMs. For constructing multiple UGMs with non-Gaussian nodes, the  only  work we could locate is \citet{Zhang2017}, where the group lasso and fused lasso penalties  are applied to construct graphs with a mixture of discrete and Gaussian nodes via a pseudo-likelihood approach.

In this paper, we aim to develop an approach that provides an unified framework to effectively jointly train multiple UGMs while satisfying the expectation on edge sparsity in each graph and dissimilarity sparsity across graphs. Toward that end, we extend the data augmentation concept of the PANDA technique developed by \citet{panda1} that regularizes single UGM estimation to jointly learning the structures of multiple graphs. Specifically, we design two types of noise to be tagged onto the observed data. The first type regularizes the estimation of each graph; and the second type  is designed to promote either the structural or the numerical similarities on the edges in the same position across multiple graphs. The noisy data matrices are first tagged onto the observed data in each graph, and then the data are combined across all graphs. Finally, GLMs are run node by node on the augmented combined data to simultaneously constructing the graphs. Our contributions  are  summarized as follows.
\begin{itemize}[leftmargin=0.2in]\setlength\itemsep{0.1em}
\item To the best of our knowledge, the extension of PANDA to the multiple graph setting, referred to as PANDAm hereafter (``m''  for ``multiple''), offers the first general framework to jointly construct multiple UGMs (GGMs included), when each node in the graphs given the other nodes can be modelled by an exponential family.
\item The augmented noisy data in PANDAm facilitate regularizing the structures of single graphs as well as imposing similarity constraints across graphs.  For the latter, we design noises to  yield  either the joint group lasso (JGL) penalty to promote structural similarities, or the joint fused ridge (JFR) penalty to promote numerical similarities on the edges in the same position across multiple graphs. Both the JGL and the JFR regularizers are convex. In addition, both regularizers can be used for variable selection and estimation in GLMs, in which the JGL is equivalent to the sparse group lasso  regularization \citep{sgl2013}, and the JFR is equivalent to ``sparse fused ridge''.
\item PANDAm recasts the constrained optimization problem  as solving maximum likelihood estimation (MLE) from GLMs iteratively from noise-augmented data. As such, it does not employ complex optimization techniques but may leverage existing software on GLMs when $\sum_{l=1}^q n^{(l)}+n_{e,1}+n_{e,2}>3p$ ($n^{(l)}$ is the sample size of the observed data in graph $l=1,\ldots,q$,  $n_{e,1}$ and $n_{e,2}$ are the sizes of the two types of augmented noises, and $p$ is the number of nodes in each graph).  The variance terms of the augmented noises are adaptive to the most up-to-date parameter estimates in each iteration. 
\item When the graphs are GGMs, besides NS, we also develop a PANDAm approach based on the  Cholesky decomposition of the precision matrices, extending \citet{Huang2006} for single GGM estimation, and a PANDAm approach that realizes regularization through quadratic optimization, extending the Sparse Column-wise Inverse Operator (SCIO) for single GGM estimation \citep{weidong2015}.
\item  The theoretical properties established for for the noise-augmented loss function by \citet{panda1} in the single graph setting also apply to the multiple graph case, including the Gaussian tail bound  and the almost sure (a.\  s.\ ) convergence  of the loss function to its expectation.
\item We offer a Bayesian interpretation for PANDAm and connect PANDAm with the empirical Bayes framework and \emph{maximum a posteriori probability} (MAP) estimate in each iteration of PANDAm.
\end{itemize}



The rest of the paper is organized as follows. Section \ref{sec:multiple} presents the PANDAm technique for joint estimation of multiple UGMs and GGMs and offers  a Bayesian perspective on PANDAm. Section \ref{sec:simulation} applies PANDAm in simulated multiple graphs and compares its performance against the commonly used estimation approaches. Section \ref{sec:case} applies PANDAm to a real-life lung cancer microarray data to construct four GGMs simultaneously. The paper concludes in Section \ref{sec:discussion} with  some final remarks.



\section{PANDAm for Simultaneous Construction of  Multiple Graphs}\label{sec:multiple}
The key component in the extension of PANDA  to the multiple graph setting is to design two type of noises that regularizes each graph and promotes structural or numerical similarities among the edges in the same position across graphs. We briefly review PANDA for single graph estimation, and its regularization effects and theoretical properties in Sec \ref{sec:pandasingle}. We propose PANDAm-JGL and PANDAm-JFR for constructing multiple GGMs and UGMs in the NS framework in Sec \ref{sec:NS}. For multiple GGMs,  we also provide the PANDAm-CD and PANDAm-SCIO approaches with the JGL and the JFR regularizations in Sec \ref{sec:CDMGGM} and \ref{sec:SCIOMGGM}, respectively. The Bayesian interpretation is provided in Section \ref{sec:EB}.


\subsection{Overview on PANDA for single graph estimation}\label{sec:pandasingle}
PANDA is a noise augmentation (NA) technique for regularizing GLM and single UGM estimation, and it belongs to the family of noise injection (NI) regularization techniques. NI may refer to adding or multiplying noises directly onto data (the data remain the original dimension $n\times p$) or attaching a noise matrix on to the observed data (either $n$ or $p$ increases). NI has been proved to be effective in promoting generalization abilities of machine learning methods, such as deep learning (see \citet{panda1} for a brief overview on NI). The model training and optimization in NI is realized through iterative algorithms with adaptive noises injected in each iteration.


PANDA offers several noise types and noise generating distributions (NGD), each leading to a different regularization effect on the edge estimation. Table \ref{tab:noise-reg} lists some examples of the  noise types  in the regression-based NS framework (see \citet{panda1} for a more complete list).  $e_{jk}$ denotes the augmented data to covariate node $X_k$ and $\theta_{jk}$ denotes the regression coefficient for $X_k$ when a GLM is run with outcome node $X_j$. If $\theta_{jk}= 0$, then there is no edge between nodes $k$ and $j$. The value of augmented noise $e_{jj}$ to outcome $X_j$ is $\bar{x}_j$, where $\bar{x}_j$ is the sample mean of the data in node $j$.  
\begin{table}[!htb]
\begin{center}
\resizebox{\columnwidth}{!}{
\begin{tabular}{lll}
\hline
noise type  & NGD ($k\neq j$) &  regularization effect $P(\Theta)$\\
\hline
bridge & $e_{jk}\sim N\left(0, \lambda|\theta_{jk}|^{-\gamma}\right)$ & $(\lambda n_e)\sum_{j=1}^{p}\sum_{k\ne j}|\theta_{jk}|^{2-\gamma}$ \\
elastic-net & $e_{jk}\sim N\left(0, \lambda|\theta_{jk}|^{-1}+\sigma^2\right)$ & $(\lambda n_e)\sum_{j=1}^{p}\sum_{k\ne j}|\theta_{jk}|+(\sigma^2 n_e)\sum_{j=1}^{p}\sum_{k\ne j}\theta_{jk}^2$ \\
adaptive lasso& $e_{jk}\sim N\left(0, \lambda |{\theta}_{jk}|^{-1}|\hat{{\theta}}_{jk}|^{-\gamma}\right)$ & $(\lambda n_e)\sum_{j=1}^{p}\sum_{k\ne j}|\theta_{jk}||\hat{\theta}_{jk}|^{-\gamma}$ \\
\hline
\end{tabular}
}
\begin{tabular}{l}
\footnotesize in the bridge type, $\gamma =1$ leads to the lasso regularizer; and $\gamma=2$ leads to the ridge  regularizer.\\
\footnotesize$\gamma\in(0,2],\lambda >0; \sigma^2>0$ are tuning parameters. $\hat{\theta}_{jk}$ in adaptive lasso is a consistent estimate for ${\theta}_{jk}$.\textcolor{white}{.\hspace{1cm}.}\\
\footnotesize PANDA also yields SCAD, group lasso, fused ridge types of regularization; see \citet{panda1}\\
\hline
\end{tabular}
\end{center}
\caption{PANDA regularization noises and the corresponding regularization effects} \label{tab:noise-reg}
\end{table}
When nodes are of different types in a graph, say $X_j$ is Gaussian while $X_k$ is Poisson, due to the asymmetry in the regression models on $X_j$ and $X_k$, $\theta_{jk}$ and $\theta_{kj}$ would have different interpretations  from a regression perspective. However,  the actual magnitude of $\theta_{jk}$ would not be important if the goal is to decide there an edge ($\theta_{jk}=0$ or not).  In the PANDA algorithm, a threshold $\tau_0$ is often prespecified for setting $\hat{\theta}_{jk}=\hat{\theta}_{jk}=0$. When $|\hat{\theta}_{jk}\hat{\theta}_{kj}|<\tau_0$, then there is no edge; otherwise, there is an edge between nodes $j$ and $k$. 
Once the noisy data are generated from a NGD, they are tagged onto the observed data and the augmented data now have a sample size of $n+n_e>p$, where $n_e$ is the  size of the noisy data. With the enlarged sample size, existing software on GLMs or regression models can be employed to get obtain the MLE on $\theta_{jk}$. The procedure iterates until convergence with the variance term in the NGD updated every iteration with the newly estimated  $\hat{\theta}_{jk}$ from the last iteration.

The regularization effects listed in Table \ref{tab:noise-reg} in obtained by taking expectation of the noise augmented loss function over the distribution of the augmented noise. In practice, the expectation can be realized by either letting $n_e\rightarrow\infty$ while keeping $\V(e_{jk})n_e\sim O(1)$ for a given $\theta_{jk}$, in each iteration, or by taking the average of noise-augmented loss functions from $m\rightarrow\infty$  consecutive iterations. 
Letting $n_e\rightarrow\infty$
or $m\rightarrow\infty$  would lead to the same targeted regularization effect for GGM. For non-Gaussian UGM,  $n_e\rightarrow\infty$ would achieve the targeted regularization effect arbitrarily well, while  $m\rightarrow\infty$ would lead to the second-order approximation of the targeted regularization effect and the higher-order deviation is negligible when $|\theta_{jk}|$ is not so large or the tuning parameters from the NGDs are small \citep{panda1}. 

For GGM, in addition to the NS approach, PANDA can estimate the GGM through the CD of the precision matrix, through the SCIO estimator of the precision matrix, through regularization of partial correlation coefficients to identify the hub nodes as in the SPACE approach \citep{Jie2009}, or through regularizing the precision matrix as a whole with the graphical ridge penalty.

\subsection{PANDAm-NS for simultaneous construction of multiple UGMs}\label{sec:NS}
Assume there are $q$ graphs with the same set of nodes, the data of which are collected under different conditions. The observed sample size $n^{(l)}$ may differ by graph for $l=1,\ldots,q$.
We propose PANDAm for the simultaneous multiple graph estimation in the NS framework by performing node-wise GLM on the combined noise-augmented data across multiple graphs, assuming that the conditional distribution of node $X_j$ ($j=1,\ldots,p$) given all other nodes $\X_{-j}$  can be modelled via an exponential family
\begin{equation}\label{eqn:expfam}
p(X^{(l)}_j|\X^{(l)}_{-j})=\exp\left(X^{(l)}_j\eta^{(l)}_j-B_j(\eta^{(l)}_j)+h_j(X^{(l)}_j)\right),
\end{equation}
where $\X^{(l)}_{-j}=(X^{(l)}_1,\ldots,X^{(l)}_{j-1},X^{(l)}_{j+1},\ldots,X^{(l)}_p)'$, and $\eta^{(l)}_j$ is the natural parameter. In the GLM framework, $\eta^{(l)}_j=\theta^{(l)}_{j0}+\sum_{k\ne j}\theta^{(l)}_{jk}X^{(l)}_k$ if the canonical link function is used  (e.g., the identity link for Gaussian $X_j$ and the logit link for Bernoulli $X_j$). If $\theta^{(l)}_{jk}=0$  or  $\theta^{(l)}_{kj}=0$, then there is no edge between nodes $j$ and $k$; otherwise, the two nodes are connected with an edge. \citet{yang2014,UGMEXP2015} prove that asymptotically with probability 1 that the neighborhood structure of conditionally exponential family graphical models can be recovered exactly.

To estimate $\bs{\theta}_{jk}=\left(\theta_{jk}^{(1)},\ldots,\theta_{jk}^{(q)}\right)$ simultaneously, we employ two types of noise. The first type, denoted by $e_1$, regularizes the estimation of each graph and can be generated from any NDG given in \citet{panda1}. That is,
\begin{align}\label{eqn:e1}
e^{(l)}_{ijk,1} \overset{\text{ind}}{\sim} N\left(0,\V(e^{(l)}_{ijk,1})\right)
\end{align}
for $i=1,\ldots, n_{e_1}$ and $k\ne j$. The actual form of $\V(e_{jik,1}^{(l)})$ depends on the targeted regularization effect in each graph (e.g.,  if the bridge-type noise is used,  then $\V(\theta_{jk}^{(l)},\bs\lambda)=\lambda^{(l)}_1|\theta^{(l)}_{jk}|^{-\gamma}$ as given in Table \ref{tab:noise-reg}, where  $\lambda^{(l)}_1$ is the graph-specific tuning parameter).  The second type of noise, denoted by $e_2$, achieves either the joint group lasso (JGL) or joint fused ridge (JFR) regularization on $\bs{\theta}_{jk}$. The JGL regularization, realized through the noise generated from Eqn (\ref{eqn:JGLUGMe2}),  promotes structural similarity among  the $q$ graphs by placing the group lasso penalty on $\bs{\theta}_{jk}$; the JFR regularization, realized through the noise generated from Eqn (\ref{eqn:JFRUGMe2}), promotes the numerical similarity by placing the fused ridge penalty $\bs{\theta}_{jk}$.
\begin{align}
\mbox{JGL: } &\textstyle \e_{ijk,2}=\left({e}^{(1)}_{ijk,2}\ldots,{e}^{(q)}_{ijk,2}\right)\overset{\text{ind}} {\sim} N\left(0,{\lambda_2}\left(\sum_{l=1}^{q}\theta_{jk}^{(l)2}\right)^{-1/2}\mathbf{I}\right)\mbox{ for $k\ne j$ } \label{eqn:JGLUGMe2} \\ 
\mbox{JFR: } & \textstyle
\e_{ijk,2}= \left({e}^{(1)}_{ijk,2}\ldots,{e}^{(q)}_{ijk,2}\right)\overset{\text{ind}} {\sim} N\left(0,\lambda_2\mathbf{T}\mathbf{T}'\right)\mbox{ for $k\ne j$},\label{eqn:JFRUGMe2} 
\end{align}
where $\mathbf{I}_{q\times q}$ is the identity matrix, the entries in matrix $\mathbf{T}$ are $T_{s,s}=1,T_{s+1-s\cdot1(s=q),s}=-1$ for  $s=1,\ldots,q$; and 0 otherwise.  The the augmented noises $e_{jj,1}$ and  $e_{jj,2}$ for outcome node $X_j$ are set at $\bar{\x}_j$, where $\bar{\x}_j$ is the sample mean of the  observed data in node $j$ combined from all graphs,
\begin{align}\label{eqn:ejj}
\textstyle e^{(l)}_{ijj,h}=\left(\sum_{l=1}^qn^{(l)}\right)^{-1}\sum_{l=1}^q\sum_{i=1}^{n^{(l)}}x_{ij}^{(l)}\mbox{ for $l=1,\ldots,q, h=1,2$ and $i=1,\ldots, n_{e_1}+n_{e_2}$}.
\end{align}
For GGM, with centered outcome nodes, $e^{(l)}_{ijj,h}=0$.
In addition to the JGL and JFR regularizers, a joint fused lasso type of regularization can be imposed $\bs{\theta}_{jk}$ by using $|\theta_{k}-\theta_{k'}|^{-1}$ ($k\ne k'$) in the covariance matrix of $\e_{ijk,2}$. However, it does not outperform the JFR penalty in terms of promoting similarity on parameter estimation (or similarity on edge patterns when jointly estimating multiple graphs). We therefore we focus the discussion on the fused ridge in the rest of the paper.

\begin{figure}[!htb]
\includegraphics[width=0.9\textwidth]{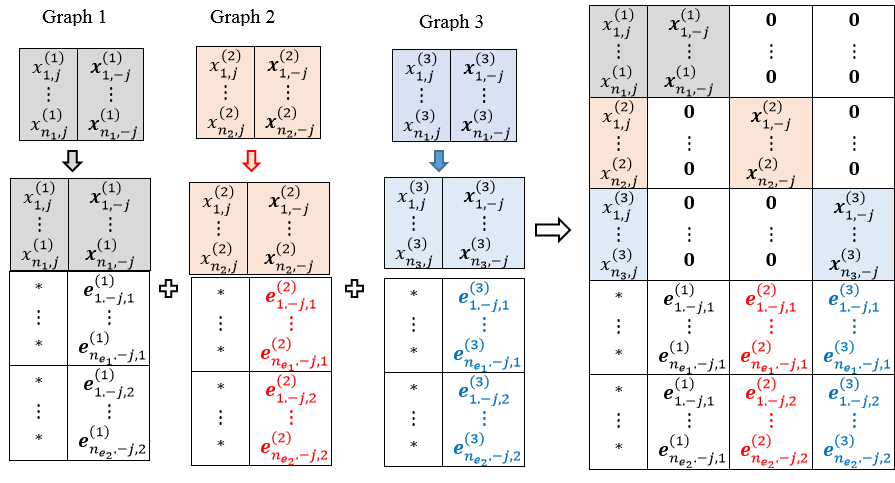}
\caption{A schematic of data augmentation for three graphs in PANDAm (noise * augmented to the outcome nodes $X_j$ is $\bar{\x}_j$, as defined in Eqn (\ref{eqn:ejj}))}\label{fig:pandaJGL}
\end{figure}

Once the noisy data $\e_1$ of size $n_{e,1}$ and $\e_2$ of $n_{e,2}$ for each graph are generated, they are tagged onto the observed data as illustrated in Figure  \ref{fig:pandaJGL}, which  uses $q=3$ as an example, but the way the data are combined is similar for all $q\ge2$).  We recommend centerizing the observed data on each ``covariate'' node in $X^{(l)}_{-j}$ and  standardizing all nodes if the graphs are GGM (or standardizing $\X^{(l)}_{-j}$ and centering the ``outcome'' node $X^{(l)}_j$) prior to the augmentation. 
The $j$-th GLM on the combined data from all $q$ graphs has node $X_j$ as the outcome and $q\times(p-1)$ covariate nodes. In the example in Figure \ref{fig:pandaJGL}, the first set $p-1$ covariate values corresponding to the first $n^{(1)}$ observations for $X_j$ comes from the first graph, while the second and third sets of  $p-1$ covariate values are all 0; similarly for the data from the second and third graphs. The noisy data are combined in a columnwise fashion rather than ``diagnonally'' as in as the observed  data. 

Let $\Theta$ denote the collection of all the regression parameters from the $p$ regression models, with  $q\times (p-1)$ regression coefficients (excluding the intercepts) per model. $\x\!=\!(\x^{(1)},\ldots,\x^{(q)})$ contains the observed data across $q$ graphs, $\e_1\!=\!(\e_1^{(1)},\ldots,\e_1^{(q)})$, and $\e_2=(\e_2^{(1)},\ldots,\e_2^{(q)})$ denote the augmented noises of types 1 and 2 across the graphs.  The summed negative log-likelihood function over $p$ GLM and $q$ graphs on the combined noise-augmented data is
\begin{align}\label{eqn:lossMUGM}
&l_p(\Theta|\x,\e_1,\e_2)=l(\Theta|\x)+ l(\Theta|,\e_1,\e_2),\mbox{ where }\\
&\textstyle l(\Theta|,\e_1,\e_2)\!=\!-\sum_{l=1}^{q}\!\sum_{j=1}^{p}\!\sum_{h=1}^2\!\sum_{i=1}^{n_{e,h}}\! \left(h_j\left(e^{(l)}_{ijj,h}\right)\!+\!\left(\!\theta_{j0}^{(l)}\!+\!\sum_{k\ne j} \left(\theta_{jk}^{(l)}e_{ijk,h}^{(l)}\!\right)\!\right)e_{ijj,h}\!-\!B_j\left(\eta_{ij,h}^{(l)}\right)\right),\notag
\end{align}
and $\eta_{ij,h}^{(l)}=\textstyle \theta_{j0}^{(l)}+\sum_{k\ne j}{\theta}_{jk}^{(l)}e_{ijk,h}^{(l)}$ when the canonical link is used. For GGM with centered nodes, we may set $\theta_{j0}^{(l)}=0$ for all $j$ and $l$. 
Proposition \ref{prop:ElossUGM} establishes the expected regularization effects of PANDAm-NS for joint estimating multiple UGMs. The proof can be found in Appendix \ref{app:prop1}.
\begin{pro}[\textbf{Regularization effects of PANDAm-NS-JGL and PANDAm-NS-JFR}]\label{prop:ElossUGM}
The expectation 
of $l_p(\Theta|\x,\e_1,\e_2)$ in Eqn (\ref{eqn:lossMUGM})  over the distribution of $\e_1$ in Eqn (\ref{eqn:e1})
and $\e_2$ in Eqns (\ref{eqn:JGLUGMe2}) or (\ref{eqn:JFRUGMe2}) is
\begin{align}
&\E_{\e_1,\e_2}(l_p(\Theta|\x,\e_1,\e_2))\!=\!
l(\Theta|\x)+P_1\left(\Theta\right)+P_2\left(\Theta\right)\!+\! C\!+\!O\!\left( \!\sum_{h=1}^2\sum_{j=1}^{p}\!\sum_{k\ne j} \!\left[n_{e,h}\theta_{jk}^4\V^2(e_{jk,h})\!\right]\!\right)\label{eqn:Ereg}\\
&\mbox{with } P_1\left(\Theta\right) =\textstyle\frac{n_{e,1}}{2}\sum_{j=1}^{p}B_j''\left(\sum_{l=1}^{q}\theta_{j0}^{(l)}\right)\sum_{l=1}^{q}\sum_{k\ne j}\theta_{jk}^{(l)2}\mbox{V}\left(e_{ijk,1}^{(l)}\right) \mbox{ and} \notag\\
&\qquad\; P_2\left(\Theta\right)\!=\!
\begin{cases}
\frac{\lambda_2 n_{e,2}}{2}\sum_{j=1}^{p}B_j''\left(\sum_{l=1}^{q}\theta_{j0}^{(l)}\right)\sum_{k\ne j}\left(\sum_{l=1}^{q}\theta_{jk}^{(l)2}\right)^{1/2} & \mbox{for JGL}\\
\frac{\lambda_2 n_{e,2}}{2}\! \sum_{j=1}^{p}\!B_j''\!\left(\!\sum_{l=1}^{q}\!\theta_{j0}^{(l)}\right)\sum_{k\ne j}\sum_{l,v\in\mathcal{S}} \!\left(\theta_{jk}^{(l)}\!-\!\theta_{jk}^{(v)}\right)^{\!2} & \mbox{for JFR}
\end{cases},\notag
\end{align}
where $\mathcal{S}$  in $P_2(\Theta)$ denotes the combinatorics set $(_2^q)$ among the $q$ graphs, and  $C$ is a known constant that relates to $n_{e,1}, n_{e,2}$ and $p$: for GGM,
$C=0$; for non-Gaussian UGM, $C=-\sum_{l=1}^{q}\sum_{j=1}^{p}\sum_{h=1}^2\sum_{i=1}^{n_{e,h}} \left(h_j\left(e_{ijj,h}\right)+\theta_{j0}^{(l)}e_{ijj,h}-B_j(\sum_{l=1}^{q}\theta_{j0}^{(l)})\right)$.
\end{pro}


Operationally, $\Theta$ are estimated through running $p$ GLM on the noised augmented data combined across the $q$ graphs in an iterative fashion. The algorithmic steps are listed in Algorithm  \ref{alg:JGL}. Most remarks on the specification of the algorithmic parameters for the PANDA algorithms in single graph estimation in \citet{panda1} apply directly to the multiple graph case, which are summarized in Remark \ref{rem:pandarem}. The only additional consideration is the relative sizes on $n_{e,1}$ and $n_{e,2}$ given in Remark \ref{rem:ne12}.
\begin{algorithm}[!htb]
\caption{PANDAm-NS-JGL and PANDAm-NS-JFG  for joint estimation of $q$ UGMs}\label{alg:JGL}
\begin{algorithmic}[1]
\State \textbf{Input}:
\begin{itemize}[leftmargin=0.18in]\setlength\itemsep{-1pt}
\item random initial estimates $\bar{\bs{\theta}}_j^{(l)(0)}$ for $j=1,\ldots,p$ and $l=1,\ldots,q$.
\item a NGD to generate $e_1$ (Eqn (\ref{eqn:e1})), a NGD to generate  $e_2$ (either Eqns (\ref{eqn:JGLUGMe2}) or (\ref{eqn:JFRUGMe2})), noisy data sizes $n_{e,1}$ and $n_{e,2}$, maximum iteration $T$, threshold $\tau_0$, width of moving average (MA) window $m$, banked parameter estimates after convergence $r$.
\end{itemize}
\State $t\leftarrow 1$; convergence $\leftarrow 0$
\State \textbf{WHILE} $t<T$ \textbf{AND} convergence $= 0$
\State \textbf{\hspace{0.3cm} FOR} $j = 1:p$
\begin{enumerate}[leftmargin=0.5in]\setlength\itemsep{-1pt}
\item[a)] Generate $e_{ijk,1}^{(l)}$  for $i=1,\ldots,n_{e,1}$, and $e_{ijk,2}^{(l)}$ for $i=1,\ldots,n_{e,2}$, with $\bar{\bs{\theta}}_j^{(l)(t-1)}$ plugged in the variance terms of the NGDs.
\item[b)] Centerize $\x^{(l)}_{-j}$ in each graph for non-Gaussian UGMs or standardize $\x^{(l)}$ for $l=1,\ldots,q$ if the graphs are GGM.
\item[c)] Obtain augmented data $\tilde{\x}$ by combining the original data $\x$ with $\e_1$ and $\e_2$ in a similar manner as in Figure \ref{fig:pandaJGL}.
\item[d)] Run GLM with outcome node $\tilde{\x}_j$ with the a proper canonical link function and linear predictor $\sum_{l=1}^q\theta^{(l)}_{j0}+\tilde{\x}^{(1)}_{-j}\bs{\theta}^{(1)}_j+\tilde{\x}^{(2)}_{-j}\bs{\theta}^{(2)}_j+\ldots+\tilde{\x}^{(q)}_{-j}\bs{\theta}^{(q)}_j$ to obtain the MLE  $\hat{\bs{\theta}}_{j}^{(l)(t)}$ for $l=1,2,\ldots,q$.
\item[e)] If $t>m$, calculate the MA $\bar{\bs{\theta}}^{(l)(t)}_j=m^{-1}\sum_{b=t-m+1}^{'} \hat{\bs{\theta}}_{j}^{(l)(b)}$; otherwise $\bar{\bs{\theta}}^{(l)(t)}_j=\hat{\bs{\theta}}^{(l)(t)}_j$ for $l=1,\ldots,q$.
\end{enumerate}
\hspace{0.45cm} \textbf{End FOR}
\State \hspace{0.45cm} Plug in $\bar{\bs{\theta}}^{(l)(t)}_j$ in Eqn (\ref{eqn:lossMUGM}) to calculate the overall loss function and apply one of the convergence
\State \hspace{0.45cm} criteria (Remark \ref{rem:pandarem}) to $\bar{l}^{(t)}$. Let convergence $\leftarrow 1 $ if the convergence is reached.
\State  \textbf{End WHILE}
\State Continue to execute the command lines 4 and 6  for another $r$ iterations, and record $\bar{\bs\theta}^{(l)(b)}_j$ for $b=t+1,\ldots,t+r$ and $l=1,\ldots,q$. Let $\bar{\bs{\theta}}^{(l)}_{jk}=(\bar{\theta}_{jk}^{(l)(t+1)},\ldots,\bar{\theta}_{jk}^{(l)(t+r)})$.
\State  Set $\hat{\theta}^{(l)}_{jk}=\hat{\theta}^{(l)}_{kj}=0$ and claim no edge between nodes $j$ and $k$
if $\left(\big|\max\{\bar{\bs{\theta}}^{(l)}_{jk}\}\cdot\min\{\bar{\bs{\theta}}^{(l)}_{jk}\}\big|<\tau_0\right) \cap \left(\max\{\bar{\bs{\theta}}^{(l)}_{jk}\}\cdot\min\{\bar{\bs{\theta}}^{(l)}_{jk}\}<0\right)$ or
$\left(\big|\max\{\bar{\bs{\theta}}^{(l)}_{kj}\}\!\cdot\!\min\{\bar{\bs{\theta}}^{(l)}_{kj}\}\big|<\tau_0\right) \cap \left(\max\{\bar{\bs{\theta}}^{(l)}_{kj}\}\!\cdot\!\min\{\bar{\bs{\theta}}^{(l)}_{kj}\}<0\right)$; otherwise, there is an edge between nodes $j$ and $k$. 
\State \textbf{Output}: $\hat{\bs{\theta}}^{(1)},\ldots,\hat{\bs{\theta}}^{(q)}$.
\end{algorithmic}
\end{algorithm}
\normalsize
\begin{rem}[\textbf{Algorithmic parameters specification}]\label{rem:pandarem}
Maximum iteration $T$ should be set at a relatively large value to ensure the algorithm meets the convergence criteria, including visual examination of the trace plot of loss function $\bar{l}^{(t)}$ to see if it stabilizes and plateaus with only mild random fluctuation, calculation of the percentage change in the loss function $|\bar{l}^{(t+1)}-\bar{l}^{(t)}|/\bar{l}^{(t)}$ to see if it falls below a prespecified threshold, or application of a formal statistical test on whether the algorithm converges (provided in Sec. 4.4 of \citet{panda1}). It should be noted, similar to the single graph estimation,  that the estimate $\hat{\theta}^{(l)}_{jk}$ cannot be exactly 0, so it is necessary to use a cutoff $\tau_0$ (often  a close-to-0 positive value),  which is justified in an asymptotic sense in that any fluctuation around $\hat{\theta}_{jk}$  diminishes as $n_e\rightarrow\infty$ or $m\rightarrow\infty$.
\end{rem}
\begin{rem}[\textbf{Choice of $n_{e,1},n_{e,2}$ and $m$}]\label{rem:ne12}
Large $n_{e,1},n_{e,2}$ or $m$ are needed to obtain the expected regularization effects. 
In practice, we would first ensure $\sum_{l=1}^q n^{(l)}+n_{e,1}+n_{e,2}>qp$ so that there are unique solution for the MLE in each regression at each iteration.  $n_{e,1}V(e_{jk,1})$ relates to the amount of regularization on single graph estimation, while  $\lambda_2n_{e,2}$ relates to the amount of regularization on edge similarity across multiple graphs.  Though $n_{e,1}$ and $n_{e,2}$ are specified independently, they are somewhat related. For example, if the lasso-type of noise is used and $\lambda_1n_{e,1}$ is large, then each individual graph would be sparse and the graphs may already be similar enough even if $\lambda_2 n_{e,2}$ is not large. If the single graphs are still dense after properly tuning $\lambda_1n_{e,1}$, and if it is believed that the dense structure would remain roughly the same across graph except for a few disagreements, then $n_{e,2}\lambda_2$ can be set a large value to promote the edge similarity.  Finally, though PANDAm allows $n_{e,1}$ and  $n_{e,2}$ vary by graph, this additional complexity and flexibility is unnecessary as the same expected regularization effects can be achieved by using the same $n_{e,1}$ and $n_{e,2}$ across graphs as long as $m$ is relatively large. 
\end{rem}
A special case of UGM is GGM where the nodes follow multivariate Gaussian distributions.  Constructing the  GGM is equivalent to estimating the precision matrix $\Omega$ of the multivariate Gaussian distribution. Specifically, if the entry $\omega_{jk}$ for $j\ne k$ is 0, then there is no edge between nodes $j$ and $k$; otherwise, there is. In the case of GGM, we would first standardize the observed data in each node of each graph, then the augmented noises $e_{ijj,1}$ and $e_{ijj,2}$ to the outcome nodes can be set at 0. Algorithm \ref{alg:JGL} runs $p$ linear regressions in each iteration to obtain the OLS estimates on $\bs{\beta}_j$, and the loss function used in the stopping criterion is the sum of squared errors (SSE) summed over the $p$ linear regressions on the noise augmented data
\begin{align}\label{eqn:lossGGM}
l_p(\Theta|\x,\e_1,\e_2)\!=\!\sum_{l=1}^{q}\!\sum_{i=1}^{n_l}\!\sum_{j=1}^{p}\!\left(\!x_{ij}^{(l)}\!-\!\sum_{k\ne j}x_{ik}^{(l)}\theta^{(l)}_{jk}\!\right)^2\!\!\!+\!\sum_{h=1}^{2}\!
\sum_{i=1}^{n_{e,h}}\!\sum_{j=1}^{p}\!\left(\!0-\sum_{l=1}^{q}\!
\sum_{k\ne j}e_{ijk,h}^{(l)}\theta_{jk}^{(l)}\!\right)^2\!\!\!.
\end{align}
The variance of the error term $\sigma_{j}^{2}$ in each regression is estimated by $\mbox{SSE}_j/\left(\sum_{l=1}^q n^{(l)}-\nu_j\right)$, where SSE$_j$ is from the regression with outcome node $j$, and $\nu_j=\mbox{trace}(\x_j(\tilde{\x}'_j\tilde{\x}_j)^{-1}\x'_j)$ is the degrees of freedom calculated as the trace of the hat matrix on the noise-augmented data $\tilde{\x}=(\x,\e)$. It is known that for GGM the following relationship exists between $\theta_{jk}$ and $\omega_{jk}$, the $[j,k]$-th entry in the precision matrix $\omega$ of a Gaussian distribution for $k\ne j$ \citep{learning631}. That is, for the $l$-th GGM,  $\theta^{(l)}_{jk}\omega^{(l)}_{jj}=-\omega^{(l)}_{jk}$, and  $\omega_{jj}^{(l)}=\left(\sigma_j^{(l)}\right)^{-2}$.  With the estimated $\hat{\sigma}_j^2$, Algorithm \ref{alg:JGL} can output the edge weights for GGM in addition to the structures of the graphs by setting
$\hat{\omega}_{jk}=-\hat{\theta}^{(l)}_{jk}\hat{\omega}^{(l)}_{jj}=-\hat{\theta}^{(l)}_{jk}\hat{\sigma}_{j}^{-2}$. The penalty terms $P_1(\Theta)$ and $P_2(\Theta)$  in  Proposition \ref{prop:ElossUGM} for the GGM case have  closed forms and are listed in Corollary \ref{cor:GGMNS}.
\begin{cor}[\textbf{Regularization effects of PANDAm-NS for multiple GGM estimation}]\label{cor:GGMNS} The two expected regularizers in Eqn (\ref{eqn:Ereg}) realized by PANDAm-NS for $q$ GGMs are
\begin{align*}
P_1(\Theta) &=\textstyle
n_{e,1}\!\sum_{l=1}^{q}\!\sum_{j=1}^{p}\!
\sum_{k\ne j}\V\left(e^{(l)}_{ijk,1}\right)\theta_{jk}^{(l)2}\notag\\
&=\textstyle
\!\sum_{l=1}^{q}\lambda_1^{(l)} n_{e,1}\sum_{j=1}^{p}\sum_{k\ne j}\left|\theta_{jk}^{(l)}\right|^{2-\gamma}\mbox{  if the bridge-type noise is used;}\notag\\
P_2(\Theta)&=
\begin{cases}
\lambda_2n_{e,2}\sum_{j=1}^{p}\sum_{k\ne j}\left(\sum_{l=1}^{q}\theta_{jk}^{(l)2}\right)^{1/2} &\mbox{ for JGL}\\
\lambda_2n_{e,2}\sum_{j=1}^{p}\sum_{k\ne j}\sum_{l,v\in\mathcal{S}} \!\left(\theta_{jk}^{(l)}\!-\!\theta_{jk}^{(v)}\right)^{\!2} & \mbox{for JFR}
\end{cases}
\end{align*}
where $\mathcal{S}$  in $P_2(\Theta)$ denotes the combinatorics set $(_2^q)$ among the $q$ graphs.
\end{cor}

\subsection{Additional PANDAm approaches for simultaneous construction of multiple GGM}\label{sec:GGM}
With the connection between the graph structure and the precision matrix for GGM, there exist additional approaches for constructing GGMs in additional to the NS approaches given in Section \ref{sec:NS}, such as the PANDAm-CD and PANDAm-SCIO approaches presented below.

\subsubsection{PANDAm-CD for multiple GGM estimation}\label{sec:CDMGGM}
The Cholesky decomposition (CD) approach refers to estimating the precision matrix through the LDL decomposition, a variant of the CD. Compared to the NS approach in Section \ref{sec:multiple}, the CD approach guarantees symmetry and positive definiteness of the estimated  precision matrix. WLOG, let $\x^{(l)}_{n\times p}\sim N_p(\mathbf{0},\Omega^{(l)})$. Apply the CD decomposition to $\Omega^{(l)}$ for $l=1,\ldots,q$ as in $\Omega^{(l)}=L^{(l)'}\!\left(\!D^{^(l)}\!\right)^{-1}\!L^{(l)}$, such that  $|\Omega^{(l)}|\!=\!\big|D^{(l)}\big|^{-1}\!=\!\prod_{j=1}^{p}\left(\sigma_{j}^{(l)}\right)^{-2}\!, D^{(l)}\!=\!\mbox{diag}\{\sigma_{1}^{(l)2},\ldots,\sigma_{p}^{(l)2}\}$, and $L^{(l)}$ is a lower uni-triangular matrix with elements $-\theta^{(l)}_{jk}$ for $j\!>\!k$, 0 for $k\!<\!j$, and 1 for $j\!=\!k$.

Similar to PANDAm-NS, PANDAm-CD estimating the edges and $q$ precision matrices via augmenting two types of noise.  The first type  $e_{1}$ is defined in the same way as Eqn (\ref{eqn:e1}). For example, if the bridge noise is used, then
\begin{align}\label{eqn:e1JGL.CD}
e^{(l)}_{ijk,1} \overset{\text{ind}} {\sim}\! N\left(0,\lambda^{(l)}_1\sigma_j^{(l)2}|\theta^{(l)}_{jk}|^{-\gamma}\right) \mbox{ and } e^{(l)}_{ijj,1} =0 \mbox{ on standardized $\x$ for }i=1,\ldots,n_{e,1}.
\end{align}
The second type $e_2$ is then drawn from either Eqns (\ref{eqn:e2JGL.CD}) or (\ref{eqn:e2JFR.CD}) to achieve the JGL and the JFR regularizations, respectively,
\begin{align}
\mbox{JGL: } &\textstyle \left({e}^{(1)}_{ijk,2}\ldots,{e}^{(q)}_{ijk,2}\right)\overset{\text{ind}} {\sim}\! N\!\left(\!0,\lambda_2\sigma_j^{(l)2}\left(\sum_{l=1}^{q}\theta_{jk}^{(l)2}\right)^{-1/2}\mathbf{I}\right)   \mbox{for }k< j; \label{eqn:e2JGL.CD}\\
 \mbox{JFR: } &\left({e}^{(1)}_{ijk,2}\ldots,{e}^{(q)}_{ijk,2}\right)\overset{\text{ind}} {\sim} N\left(0,\lambda_2\sigma_j^{(l)2}\mathbf{T}\mathbf{T}'\right)\mbox{ for } k< j \label{eqn:e2JFR.CD}
\end{align}
for  $i=1.\ldots,n_{e,2}$, where  $\mathbf{I}_{q\times q}$ is the identity matrix, and the entries in matrix $\mathbf{T}_{q\times q}$ are $T_{s,s}=1,T_{s+1-s\cdot1(s=q),s}=-1$ for  $s=1,\ldots,q$; and 0 otherwise. $e^{(l)}_{ijj,2}$ can be set at 0 on standardized $\x^{(l)}$ for $j=1,\ldots,p$ a. 

\begin{figure}[!htb]
\includegraphics[width=1.0\textwidth]{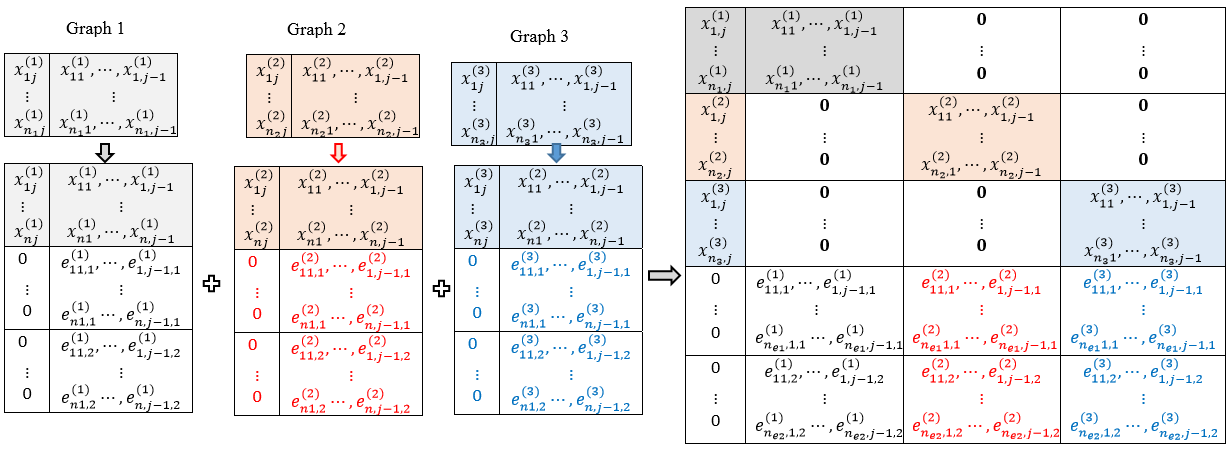}
\caption{A schematic of data augmentation in PANDAm-CD for three GGMs any $j\in[1,p]$. When $j=1$, there are no covariate nodes in the regression model, and  $\tilde{\x}_1$ (the first column in the combined augmented data) is used to estimate $\sigma_1^2$.} \label{fig:pandaCD_JGL}
\end{figure}

Once $\e_1$ and $\e_2$ are generated, they are tagged onto the original $\x$ and then the augmented data across $q$ graphs are combined, as depicted in Figure \ref{fig:pandaCD_JGL} for any $j$ from 1 to $p$. Though the data augmentation scheme looks similar to  Figure \ref{fig:pandaJGL} in the NS setting, there is an important difference, that is, the dimension of the covariate nodes grows with $j$ given how the precision matrix is estimated through the CD approach.
Specifically. Let $\bs{\epsilon}=(\epsilon_1,
\ldots,\epsilon_p)' =L^{(l)}X^{(l)}$. With a bit algebra, we get  
\begin{align}\label{eqn:CDregl}
X^{(l)}_1=\epsilon_1\mbox{ and }
X^{(l)}_j=\textstyle \sum_{k=1}^{j-1}
X^{(l)}_k\theta^{(l)}_{jk}+ \epsilon_j \mbox{ for $j=2,\ldots,p$, where $ \epsilon_j\sim  N(0,\sigma_j^{(l)2})$}.
\end{align}
With the augmented data from the graphs are combined in the way as depicted in Figure \ref{fig:pandaCD_JGL}, we update Eqn (\ref{eqn:CDregl}) to
\begin{align}\label{eqn:CDreg}
\tilde{X}_1=\tilde{\epsilon}_1\mbox{ and }
\tilde{X}_j=\textstyle\sum_{l=1}^q\sum_{k=1}^{j-1}
\tilde{X}_k\theta^{(l)}_{jk}+ \tilde{\epsilon}_j \mbox{ for $j=2,\ldots,p$, where $ \epsilon_j\sim  N(0,\sigma_j^{2})$}.
\end{align}
Eqn (\ref{eqn:CDreg}) suggests a sequence of regression models: $\tilde{X}_2$ regressed on $\tilde{X}_1$, $\tilde{X}_3$ regressed on $(\tilde{X}_1,\tilde{X}_2)$, and so on (there is no regression model for $j=1$). After solving for $\hat{\theta}^{(l)}_{jk}$ from the series of regression models, $\sigma^{(l)2}_j$ can be calculated as
\begin{align}
\hat{\sigma}^{(l)2}_j=\textstyle \left(n^{(l)}\right)^{-1}\sum_{i=1}^{n^{(l)}}\left(\x^{(l)}_{ij}-\sum_{k=1}^{j-1}\x^{(l)}_{ik}\hat{\theta}^{(l)}_{jk}\right)^2.\label{eqn:CDsigma2}
\end{align}
When $j=1$, $\hat{\sigma}^{(l)2}_1$ can be set at the sample variance of $\x^{(l)}_1$. Proposition \ref{prop:CD} show the PANDAm-CD approach described above achieves the targeted regularization effects over the distribution of $\e_1$ and $\e_2$ in the multiple GGM estimation. 
\begin{pro}[\textbf{Regularization effects of PANDAm-CD for multiple GGM estimation}]\label{prop:CD}
The noise-augmented loss function across the $q$ graphs is
$$l_p(\Theta|\x,\!\e_1,\!\e_2\!)\!=\!\sum_{l=1}^{q}\!\sum_{i=1}^{n_l}\!\sum_{j=1}^{p}\!\left(\!\sigma_{j}^{(l)}\!\right)^{\!-2}\!\!\left(\!x_{ij}^{(l)}\!-\!\sum_{k=1}^{j-1}x_{ik}^{(l)}\theta_{jk}^{(l)}\!\right)^2\!+\!\sum_{h=1}^{2}\!\sum_{i=1}^{n_{e,h}}\!\sum_{j=1}^{p}\!\left(\!\sigma_{j}^{(l)}\!\right)^{\!-2}\!\!\left(\!e_{ijj,h}\!-\!\sum_{l=1}^{q}\!\sum_{k=1}^{j-1}\!e_{ijk,h}^{(l)}\theta_{jk}^{(l)}\!\right)^2,$$
the expectation of which over the distribution of $\e_1$ and $\e_2$ is $\E\left(l_p(\Theta|\x,\e_1,\e_2)\right)=$
\begin{align}
\!\!\mbox{JGL: }& \textstyle l(\Theta|\x)\!+\!n_{e,1}\sum_{l=1}^{q}\!\sum_{j=1}^{p}\!\sum_{k=1}^{j-1}\!\V\left(e_{ijk,1}^{(l)}\right)\theta_{jk}^{(l)2}\!+\!\lambda_2n_{e,2}\!\sum_{j=1}^{p}\!\sum_{k=1}^{j-1}\!\left(\sum_{l=1}^{q}\theta_{jk}^{(l)2}\right)^{1/2}\label{eqn:Ecdmultigrouplasso}\\
\!\!\mbox{JFR: } &\textstyle l(\Theta|\x)\!+\!n_{e,1}\sum_{l=1}^{q}\!\sum_{j=1}^{p}\!\sum_{k=1}^{j-1}\!\V\left(e_{ijk,1}^{(l)}\right)\theta_{jk}^{(l)2}+\lambda_2 n_{e,2}\! \sum_{j=1}^{p}\sum_{k\ne j}\sum_{l,v\in\mathcal{S}} \!\left(\theta_{jk}^{(l)}\!-\!\theta_{jk}^{(v)}\right)^{\!2}\!,\!\label{eqn:Ecdmultifusedridge}\\
& \mbox{ where $\mathcal{S}$  denotes the combinatorics set $(_2^q)$ among the $q$ graphs}\notag
\end{align}
\end{pro}
If the bridge-type noise is used on $\e_1$, then the penalty term  $\sum_{l=1}^{q}n_{e,1}\sum_{j=1}^{p}\!\sum_{k=1}^{j-1}\!\V\left(e_{ijk,1}^{(l)}\right)\theta_{jk}^{(l)2}$  in Eqns (\ref{eqn:Ecdmultigrouplasso}) and (\ref{eqn:Ecdmultifusedridge}) on each graph becomes $\sum_{l=1}^{q}\lambda_1^{(l)}n_{e,1}\sum_{j=1}^{p}\!\sum_{k=1}^{j-1}\!\left|\theta_{jk}^{(l)}\right|^{2-\gamma}$, which is the  regularization \citet{Huang2006} originally proposed to estimate the precision matrix of  a single GGM by solving the constrained optimization problem $\hat{\theta}_{jk}=\textstyle \arg\min\limits_{\theta_{jk}}\hat{\sigma}_j^{-2}\sum_{i=1}^{n}\left(\x_{ij}-\sum_{k=1}^{j-1}\x_{ik}\theta_{jk}\right)^2+\xi\sum_{k=1}^{j-1}|\theta_{jk}|^\gamma\label{eqn:CDtheta}$ for $k= 1,\ldots, j-1,j+1,\ldots,p$.

It is not surprising that the penalty  forms  in Proposition \ref{prop:CD} look similar to those in Corollary \ref{cor:GGMNS} given that both are based on the linear regression framework. However, the parameter $\theta^{(l)}_{jk}$ has different meaning and interpretation for the NS and CD settings, though the notation is the same. In the NS setting, the precision matrix $\Omega$ is solved from $\theta^{(l)}_{jk}$ through  $\omega_{jj}^{(l)}=\left(\sigma_j^{(l)}\right)^{-2}$ and $\omega^{(l)}_{jk}=-\theta^{(l)}_{jk}\omega^{(l)}_{jj}$ whereas in the CD setting, $\theta^{(l)}_{jk}$ makes the lower uni-triangle matrix $L^{(l)}$ and the precision matrix is calculated as $\Omega= L^{(l)'}\diag{\left\{\left(\sigma_1^{(l)}\right)^{-2},\ldots,\left(\sigma_p^{(l)}\right)^{-2}\right\}}L^{(l)}$.

The algorithmic steps for implementing PANDAm-CD are available in Algorithm \ref{alg:CDJGLJFR} in the Supplementary Materials. In general, the steps are similar to Algorithm \ref{alg:JGL}, except that the set of covariates grows with $j$ in each iteration, and an additional inner loop is added within each regression in each iteration to alternatively solve for $\theta^{(l)}_{jk}$ and $\sigma^{(l)2}_j$ as they are inter-dependent (the solution $\hat{\sigma}^{(l)2}_j$ depends on the estimate $\hat{\theta}^{(l)}_{jk}$, which in turn depends on $\hat{\sigma}^{(l)2}_j$ as the distribution of the augmented noise based on which $\hat{\theta}^{(l)}_{jk}$ is calculated depends on $\hat{\sigma}^{(l)2}_j$).

\subsubsection{PANDAm-SCIO for multiple GGM estimation}\label{sec:SCIOMGGM}
Let  $\bs{\theta}_{j}$ be the $j$-th column of the precision matrix $\Omega_{p\times p}$ in a GGM. \citet{weidong2015} propose the SCIO estimator as the solution to a quadratic optimization problem that does not involve the maximization of the likelihood per se: $\hat{\bs{\theta}}_{j}=\textstyle \arg\min\limits_{\bs{\theta}_{j}}\left\{(2n)^{-1}\bs{\theta}_{j}^{'}\x^{'}\x \bs{\theta}_{j}-\mathbf{1}_j\bs{\theta}_{j}+\lambda\sum_{k\ne j}|{\theta}_{jk}|\right\}$ separately for $j=1,\ldots,p$. \citet{panda1} provide a PANDA counterpart to SCIO in the single graph setting and simplify the problem to calculating $\hat{\bs{\theta}}_j=n\left(\tilde{\x}^{'}\tilde{\x}\right)^{-1} \1_j,$ where $\tilde{\x}$ is the noise-augmented data and $\1_j$ is the $j^{th}$ column of the identity matrix $\mathbf{I}_p\times p$.

We now extend PANDA-SCIO for single GGM estimation to multiple GGMs with either the JGL or the JFR regularization.  The first type of noise $e_1$ is drawn from the  NGD in Eqn (\ref{eqn:expfam}). For example, if the bridge-type of noise is used, then
\begin{equation}\label{eqn:e1JGL.SCIO}
e^{(l)}_{ijk,1} \overset{\text{ind}} {\sim}\! N\left(0,\lambda^{(l)}_1\left|\theta^{(l)}_{jk}\right|^{-\gamma}\right); \mbox{ and } e^{(l)}_{ijj,1} =0 \mbox{ for }i=1,\ldots,n_{e,1}.
\end{equation}
The similarity-promoting noise $e_2$ is drawn from Eqn (\ref{eqn:e2JGL.SCIO}) and (\ref{eqn:e2JFR.SCIO}) for the JGL and JFR regularizations, respectively
\begin{align}
\mbox{JGL: } &\textstyle \left({e}^{(1)}_{ijk,2}\ldots,{e}^{(q)}_{ijk,2}\right)\overset{\text{ind}} {\sim}\! N\!\left(\!0,\lambda_2\sigma_j^{(l)2}\left(\sum_{l=1}^{q}\theta_{jk}^{(l)2}\right)^{-1/2}\mathbf{I}\right)   \mbox{for }k\neq j; \mbox{ and }e^{(l)}_{ijj,2} =0.\label{eqn:e2JGL.SCIO}\\
\mbox{JFR: } &({e}^{(1)}_{ijk,2}\ldots,{e}^{(q)}_{ijk,2})\overset{\text{ind}} {\sim} N\left(0,\lambda_2(\mathbf{T}\mathbf{T}')\right)\mbox{ for } k\neq j \mbox{; and } e_{ijj,2}=0\label{eqn:e2JFR.SCIO}
\end{align}
for $i=1,\ldots, n_{e,2}$, where $\mathbf{I}_{q\times q}$ is the identity matrix, and the entries in matrix $\mathbf{T}_{q\times q}$ are $T_{s,s}=1,T_{s+1-s\cdot1(s=q),s}=-1$ for  $s=1,\ldots,q$, and 0 otherwise. The noisy data generated from Eqns (\ref{eqn:e1JGL.SCIO}), (\ref{eqn:e2JGL.SCIO}) and (\ref{eqn:e2JFR.SCIO}) are first scaled to obtain $\sqrt{2n}\e_{j,1}^{(l)}$ and $\sqrt{2n}\e_{j,2}^{(l)}$, before being tagged onto the observed data $\x$ in a similar fashion as in Figure \ref{fig:pandaJGL}. The  algorithmic steps for carrying out PANDAm-SCIO is given in Algorithm \ref{alg:SCIO} in the supplementary materials.  Proposition \ref{prop:SCIO} establishes the regularization effects on $\Omega$ with the proof given in Appendix \ref{app:mcolumnwiseSCIO}.
\begin{pro}[\textbf{Regularization effects of PANDAm-SCIO for multiple GGMs}]\label{prop:SCIO}
Let $[*//*]$ denote two matrices combined by row and $[*||*]$ denote two matrices combined by column. Denote the collections of parameters and noises for column $j$ ($j=1,\ldots,p$) by $\Theta_j=\left[\bs\theta_j^{(1)}//\ldots//\bs\theta_j^{(q)}\right],\e_{j,1}=\left[\e_{j,1}^{(1)}||\ldots||\e_{j,1}^{(q)}\right]$, and $\e_{j,2}=\left[\e_{j,2}^{(1)}||\ldots||\e_{j,2}^{(q)}\right]$. Denote the observed data, the collections of all model parameters, and all augmented noises by $\x=\left(\x^{(1)}||\ldots||\x^{(q)}\right)$, $\Theta=\left(\Theta_1||\ldots||\Theta_p\right)$, $\e_{1}=\left(\e_{1,1}||\ldots||\e_{p,1}\right)$, and $\e_{2}=\left(\e_{1,2}||\ldots||\e_{p,2}\right)$. Let $\Xi_j=\left[\1_j//\ldots//\1_j\right]$. The noise-augmented loss function and its expectation over the distribution of $\e$ are respectively
\begin{align}
&l_p(\Theta|\x,\e_1,\e_2)=\textstyle(2n)^{-1}\sum_{j=1}^{p}\Theta_j^{'}(\tilde{\x}_j^{'}\tilde{\x}_j)\Theta_j-\sum_{j=1}^{p}\Theta_j^{'}\Xi_j\label{eqn:SCIOpelfJGL}\\
&\E(l_p(\Theta|\x,\e_1,\e_2))\label{eqn:SCIOnmpelfJGL}\\
&=\textstyle l(\Theta|\x)\!+\!\lambda_1^{(l)} n_{e,1}\!\!\sum_{j=1}^{p}\!
\sum_{k\ne j}\!\sum_{l=1}^{q}\left|\theta_{jk}^{(l)}\right|^{2-\gamma}+\begin{cases}
\lambda_2n_{e,2}\sum_{j=1}^{p}
\!\sum_{k\ne j}\left(\sum_{l=1}^{q}\theta_{jk}^{(l)2}\right)^{1/2} & \mbox{for JGL}\\
\lambda_2{n_e,2}\! \sum_{j=1}^{p}\sum_{k\ne j}\sum_{l,v\in\mathcal{S}} \!\left(\theta_{jk}^{(l)}\!-\!\theta_{jk}^{(v)}\right)^{\!2} & \mbox{for JFR}
\end{cases}\notag
\end{align}
where $\mathcal{S}$  denotes the combinatorics set $(_2^q)$ among the $q$ graphs.
\end{pro}
The minimizer of Eqn (\ref{eqn:SCIOpelfJGL}) can be solved by taking its derivative and setting it to be $0$, resulting in $\hat{\Theta}_j=n\left(\tilde{\x}_j^{'}\tilde{\x}_j\right)^{-1}\Xi_j$. Note the inverse of $\left(\tilde{\x}_j^{'}\tilde{\x}_j\right)$ exists since the augmented data $\tilde{\x}_j$ has a sample size larger than its feature dimensionality. Also noted is that $\E_\e\left(\tilde{\x}^{'}_j\tilde{\x}_j\right)$  for the JGL case is a block-diagonal matrix given that the noises in different graphs are independent, further simplifying the solution to $\hat{\Theta}_j^{(l)}=2n\left(\tilde{\x}_j^{(l)'}\tilde{\x}_j^{(l)}\right)^{-1}\1_j,l=1,\ldots,q$.

\subsection{PANDA-JGL and PANDA-JFR beyond graphical models}\label{sec:sfr}
The JGL and JFR regularizers realized through PANDAm can be applied beyond the framework of graphical models for variable selection and parameter estimation in regression models in general. Suppose there are $p$ predictors belonging to $q$ groups in a regression model (e.g., several genes on the same pathway in a regulatory network). Denote the  groups by $l=1,\ldots,q$ and the variables in group $l$ by $X_{j(l)}$ for $j=1,....p_l$.

If the goal is to introduce sparsity across variables within each group and as well at the group level across groups, then we could employ the following two types of noises  through PANDAm
\begin{align}
\mbox{lasso noise: } & e_{ij(l),1} \textstyle \overset{\text{ind}} {\sim}\! N\left(0,\lambda^{(l)}_1|\theta_{j(l)}|^{-1}\right)\mbox{ for } i=1,\ldots, n_{e_1}; j=1,\ldots,p_l; \mbox{ and } l=1,\ldots,q\label{eqn:sgl1}\\
\mbox{JGL noise: } & \left(e_{i1(l),2}\ldots,{e}_{ip_l(l),2}\right)\textstyle \overset{\text{ind}} {\sim}\! N\!\left(\!0,\lambda_2\!\left(\sum_{j=1}^{p^{(l)}}\theta_{j(l)}^{2}\right)^{-\frac{1}{2}}
\!\mathbf{I}\!\right) \mbox{ for } i
\!=\!1,\ldots, n_{e_2}; l\!=\!1,\ldots,q, \label{eqn:sgl2}
\end{align}
where $\mathbf{I}_{p_l\times p_l}$ is the identity matrix.  $e_1$ leads to sparsity across the variables within each group, while $e_2$ leads to sparsity across the groups.  The combined regularization in Eqns (\ref{eqn:sgl1}) and (\ref{eqn:sgl2}) as a matter of fact corresponds to the sparse group lasso penalty \citep{sgl2013}.

If the goal is to introduce sparsity across variables within each group while shrinking the parameters at the group level (i.e., each group as a whole), then we could employ the following two types of noises through PANDAm
\begin{align}
\mbox{lasso noise: } &  e_{ij(l),1} \overset{\text{ind}} {\sim}\! N\left(0,\lambda^{(l)}_1|\theta_{j(l)}|^{-1}\right) \mbox{ for } i=1,\ldots, n_{e_1}; j=1,\ldots,p_l; \mbox{ and } l=1,\ldots,q\label{eqn:sfr1}\\
\mbox{JFR noise: } & \left(e_{i1(l),2}\ldots,{e}_{iq_l(l),2}\right)\overset{\text{ind}} {\sim} N\left(0,\lambda_2\mathbf{T}\mathbf{T}'\right) \mbox{ for } i=1,\ldots, n_{e_2}; l=1,\ldots,q,\label{eqn:sfr2}
\end{align}
where the entries in matrix $\mathbf{T}_{p_l\times p_l}$ are $T_{jj}=1,T_{j+1-j\cdot1(j=p_l),j}=-1$ for  $j=1,\ldots,p_l$; and 0 otherwise.
To the best of our knowledge, there does not seem to exist a counterpart in the regression setting to the combined regularization introduced through the noises in Eqns (\ref{eqn:sfr1}) and (\ref{eqn:sfr2}). We hereby coin the term \emph{sparse fused ridge} regularization to refer to the  regularization yielded by Eqns (\ref{eqn:sfr1}) and (\ref{eqn:sfr2}) that can be employed for regression models. 


\subsection{Bayesian perspectives on PANDAm}\label{sec:EB}
Both the PANDA technique and Bayesian modeling impose regularization through introducing endogenous information. \citet{panda1} draw connections between the two paradigms in two aspects. First,  the various regularization effects realized by PANDA have counterpart priors on the model parameters through Bayesian hierarchical modeling. Second, the parameters estimates obtained by PANDA at every iterations can be seen as obtaining the maximum a posterior (MAP) estimate in an empirical-Bayes (EB)-like framework. 

In the PANDAm setting, there also exist priors that lead to the JGL and JFR regularizations in the Bayesian hierarchical setting. As mentioned in Sec \ref{sec:sfr}, the JFR regularization per regression is equivalent to the sparse group lasso in expectation over the distribution of augmented $e_1$ and $e_2$. Since the focus on the priors that lead to the second regularization term (JGL or JFR), we use the lasso-type $e_1$ for demonstration WLOG; the priors corresponding to the regularization effects generated by other types of $e_1$ (e.g. ridge, elastic net) can be found in the literature. 
Let $\bs\theta'_{jk}=(\theta_{jk}^{(1)},\ldots,\theta_{jk}^{(q)})$. \citet{BayesSGL} develop the Bayesian version for the sparse group lasso, which, reformulated in the context of the regression run by PANDAm, is
\begin{align}
&\pi\left(\bs\theta'_{jk}|\bs{\tau}^2,\gamma^2,\sigma_j^2\right)\!= \!N(0,\sigma_j^2\mathbf{V}),\mbox{where } \bs{\tau^2}\!=\!\left(\tau^{2(1)},\ldots,\tau^{2(q)}\right),\mathbf{V}\!=\!\diag\left\{\left[\left(\bs\tau\right)^{-2}\!+\!\gamma^{-2}\right]^{-1}\right\}\label{eqn:jgl.prior1}\\
&\pi(\bs{\tau}^{2},\gamma^2)
\propto\gamma^{-1} \prod_{l=1}^q\!\left\{\!\left(1\!+\!\frac{\tau^{2(l)}}{\gamma^2}\right)^{-\frac{1}{2}}\!\right\}
\exp\!\left\{-\frac{\lambda^{(l)2}_1}{2}\sum_{l=1}^g\!\tau^{(l)2}-\frac{\lambda^2_2}{2}\gamma^2\right\} \label{eqn:jgl.prior2}\\
&\pi(\sigma_{j}^{2})\propto\sigma_{j}^{-2},\notag
\end{align}
The priors that lead to the JFR regularization with the lasso regularization on each graph is
\begin{align}
&\pi\left(\bs\theta'_{jk}|\bs\nu,\sigma_j^2\right)\propto \exp\left\{-\frac{1}{2}\sigma_j^2\left(\bs\theta'_{jk}\left[\diag\{\bs\nu\}^{-2}+\lambda_2\mathbf{T}\mathbf{T}'\right]\bs\theta_{jk}\right)\right\}\mbox{ where } \bs\nu=(\nu^{(1)},\ldots,\nu^{(q)}) \label{eqn:jfr.prior1}\\
&\pi\left(\bs\nu=(\nu^{(1)},\ldots,\nu^{(q)})\right)=\textstyle \prod_{l=1}^q \lambda_1^{(l)2}\nu^{(l)}\exp\left(-\lambda_1^{(l)2}\nu^{(l)2}/2 \right),\label{eqn:jfr.prior2}\\
&\pi(\sigma_{j}^{2})\propto\sigma_{j}^{-2}.\notag
\end{align}
where the entries $T_{s,s}=1,T_{s+1-s\cdot1(k=q),s}=-1$ for  $s=1,\ldots,j-1, j+1,\ldots,q$; and 0 otherwise.
The $\lambda_2\bs\theta'_{jk}\mathbf{T}\mathbf{T}'\bs\theta_{jk}$ component in Eqn (\ref{eqn:jfr.prior1}) promotes the  edge  similarity numerically across graphs, whereas the mixture of the Gaussian component $\bs\theta'_{jk}\diag\left(\nu^{(1)},\ldots,\nu^{(q)} \right)^{-2}\bs\theta_{jk}$ in Eqn (\ref{eqn:jfr.prior1}) with the Rayleigh distribution on $\bs\nu$ in Eqn (\ref{eqn:jfr.prior2}) leads to a Laplace prior $\theta_{jk}$ that promotes the sparsity within each graph. 

We also provide an EB-like interpretation for PANDAm-NS in jointly constructing multiple graphs. We use the bridge-type noise for $e_1$ for illustration purposes, which can be easily modified to accommodate other NGDs.
The EB-like priors on $\theta^{(l)}_{jk}$ ($l=1,\ldots,q$ and $j=1,\ldots,k\ne j=1,\ldots,p$) in iteration $(t+1)$ that eventually lead to the JGL and JFR penalties on the edges upon convergence are
\begin{align}
\mbox{GGM JGL: }& \pi\left(\theta_{jk}^{(1)},\ldots,\theta_{jk}^{(q)}|\sigma_{j}^{2}\right)\propto \exp\left\{-\left(2\sigma_j^{2}\right)^{-1}\sum_{l=1}^q\left(
\lambda_1^{(l)}\frac{\theta^{(l)2}_{jk}}{|\hat{\theta}^{(l)(t)}_{jk}|^\gamma}+
\lambda_2\frac{\theta^{(l)2}_{jk}}{\big\|\hat{\bs\theta}^{(t)}_{jk}\big\|_2}\right)\right\}\label{eqn:prior.GGM.JGL}\\
\mbox{GGM JFR: }& \pi\!\left(\theta_{jk}^{(1)},\ldots,\theta_{jk}^{(q)}|\sigma_{j}^{2}\right)\propto \exp\!\left\{\!-\!\left(2\sigma_j^{2}\right)^{-1}\!\!\left(\sum_{l=1}^q\lambda_1^{(l)}\frac{\theta^{(l)2}_{jk}}{|\hat{\theta}^{(l)(t)}_{jk}|^\gamma}\!+\!\lambda_2\bs\theta^{'}_{jk}\!\mathbf{T}\mathbf{T}'\bs\theta_{jk}\!\right)\!\right\}\! \label{eqn:prior.GGM.JFR}\\
&\mbox{and }\pi(\sigma_{j}^{2})\propto \sigma_{j}^{-2}\mbox{ in Eqns (\ref{eqn:prior.GGM.JGL}) and (\ref{eqn:prior.GGM.JFR})};\notag\\
\mbox{UGM JGL: }&\pi\left(\theta_{jk}^{(1)},\ldots,\theta_{jk}^{(q)}\right)\propto \exp\left\{-\frac{1}{2}\sum_{l=1}^q\left(
\lambda_1^{(l)}\frac{\theta^{(l)2}_{jk}}{|\hat{\theta}^{(l)(t)}_{jk}|^\gamma}+
\lambda_2\frac{\theta^{(l)2}_{jk}}{\big\|\hat{\bs\theta}^{(t)}_{jk}\big\|_2}\right)\right\}\label{eqn:prior.UGM.JGL}\\
\mbox{UGM JFR: }& \pi\left(\theta_{jk}^{(1)},\ldots,\theta_{jk}^{(q)}\right)\propto \exp\left\{-\frac{1}{2}\left(\sum_{l=1}^q\lambda_1^{(l)}\frac{\theta^{(l)2}_{jk}}{|\hat{\theta}^{(l)(t)}_{jk}|^\gamma}+ \lambda_2\bs\theta^{'}_{jk}\mathbf{T}\mathbf{T}'\bs\theta_{jk}\right)\right\} \label{eqn:prior.UGM.JFR}
\end{align}
for GGM and UGM, respectively. Each prior above contains two additive components. The first component $\sum_{l=1}^q\lambda_1^{(l)}\frac{\theta^{(l)2}_{jk}}{|\hat{\theta}^{(l)(t)}_{jk}|^\gamma}$  regularizes the sparsity in each graph; and the second component imposes either the JGL or the JFR regularizations. Both the first component as well as the JGL prior are formulated with an estimate $\hat{\theta}^{(l)(t)}_{jk}$ from the last iteration $t$, but not the JFR prior.  

The priors listed Eqns (\ref{eqn:prior.GGM.JGL}) and  (\ref{eqn:prior.UGM.JFR}) in iteration $t+1$ are formulated using the most up-to-date parameter estimates from the augmented data in the last iteration $t$, thus the connection with EB. The parameter estimate $\hat{\theta}_{jk}^{(t)}$ can be either the MAP estimate or a random posterior sample for ${\theta}_{jk}$ from the $t$-th iteration. As pointed out in \citet{panda1}, when the expected regularization effect is convex in ${\theta}_{jk}$, which is the case for JGL, the MAPs obtained upon the convergence of the EB-like PANDAm algorithm are the same regardless of whether the prior in each iteration is constructed using the MAP or a random posterior sample. If the expected regularization effect is non-convex in ${\theta}_{jk}$, which is the case for JFR, plugging in the MAP estimate or a random posterior sample from the last iteration  for $\hat{\theta}_{jk}^{(t)}$ in the priors would lead to different MAPs upon the convergence. Using a random posterior sample would eventually lead to a  converged multi-modal posterior distribution on ${\theta}_{jk}$, while using the MAP from the last iteration for $\hat{\theta}_{jk}^{(t)}$ leads to the MAP that is an initial-value-dependent local optimum, upon convergence. This is because using MAP to construct the prior in each iteration is equivalent to iteratively solves convex optimization in the same way as PANDA, but only in the Bayesian framework while by using a random posterior draw allows the posterior sampling to explore the whole parameter space.

\section{Simulation}\label{sec:simulation}
We apply PANDAm-NS-JGL and PANDAm-NS-JFR to jointly train three GGM graphs and three PGM graphs, respectively. In the GGM case, we compare PANDA to the joint graphical lasso approach  with the group lasso (GL) penalty and the fused lasso (FL) penalty \citep{JGL, Danaher2014}, and to the sparse group lasso   with the lasso and GL penalty. Note the sparse group lasso originally is not developed  for constructing multiple graphs, but rather being used in the regression setting. We adapt the sparse group lasso to construct multiple graphs, and expect it yields similar results to PANDAm-NS-JGL.  Since there does not exist any approach for jointly estimating multiple UGMs when the  all nodes  are non-Gaussian, to the best of our knowledge, we thus compare PANDAm with the  na\"ive differencing approach in the non-Gaussian UGM case that constructs each graph separately and then compares the edge connection patterns in the estimated graphs pairwise in a post-hoc manner. 

\begin{figure}[!htb]
\begin{center}
\textcolor{magenta}{scale-free: } \begin{minipage}{0.25\textwidth}
\includegraphics[width=1\linewidth]{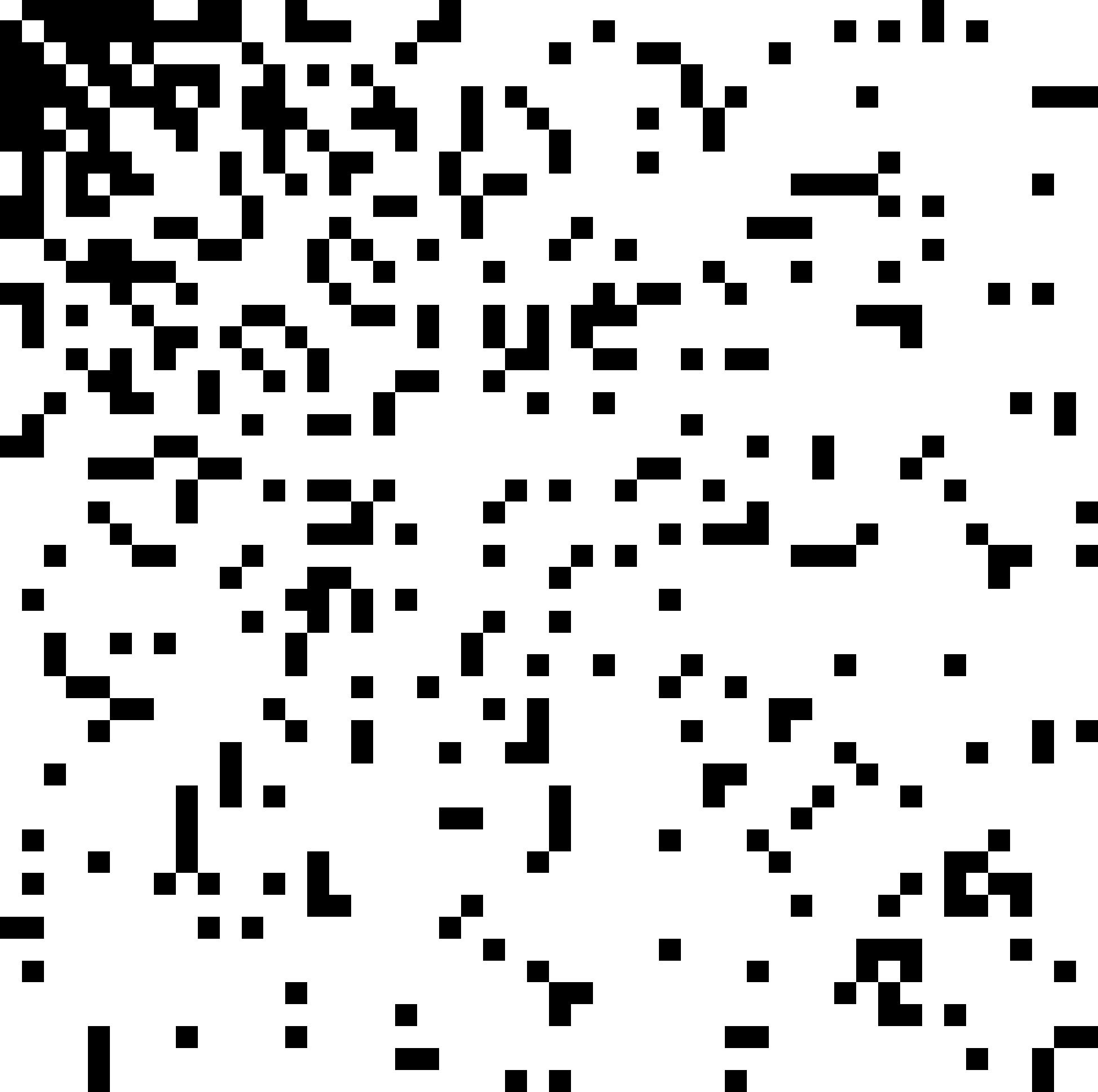}
\end{minipage}
\begin{minipage}{0.25\textwidth}
\includegraphics[width=1\linewidth]{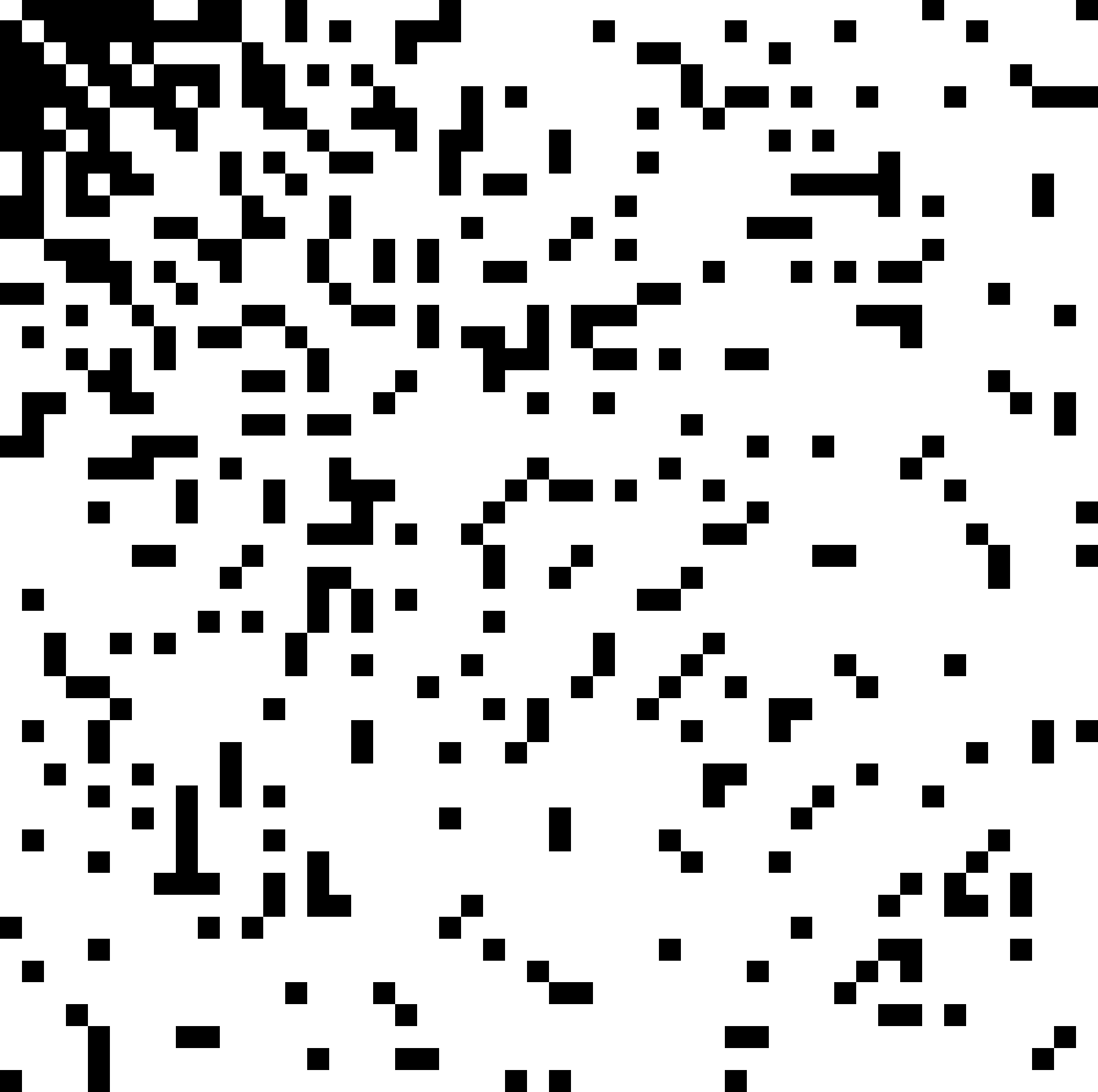}
\end{minipage}
\begin{minipage}{0.25\textwidth}
\includegraphics[width=1\linewidth]{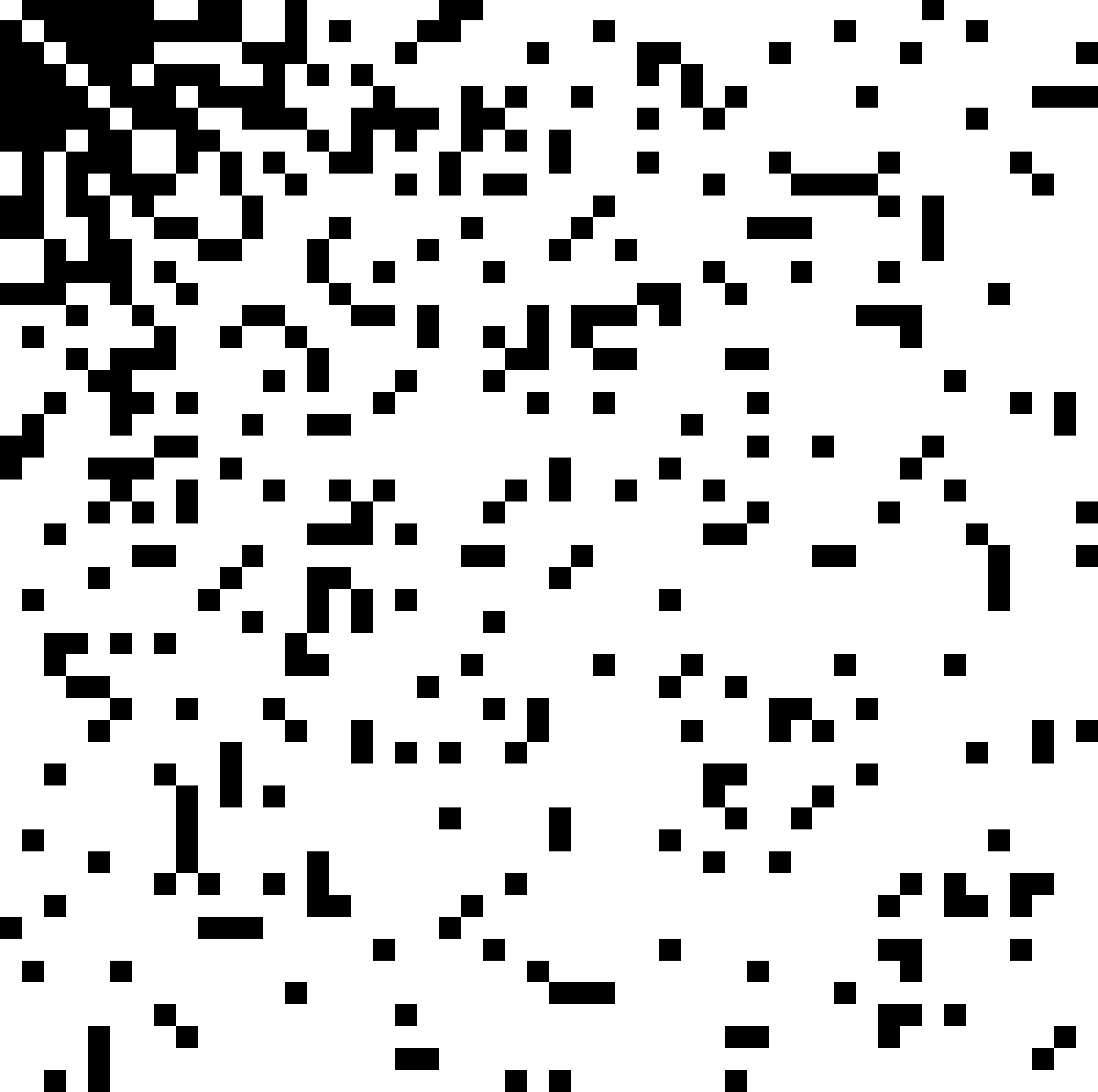}
\end{minipage}\\
\textcolor{magenta}{banded $\mathbf{A}$: }
\begin{minipage}{0.25\textwidth}
\includegraphics[width=1\linewidth]{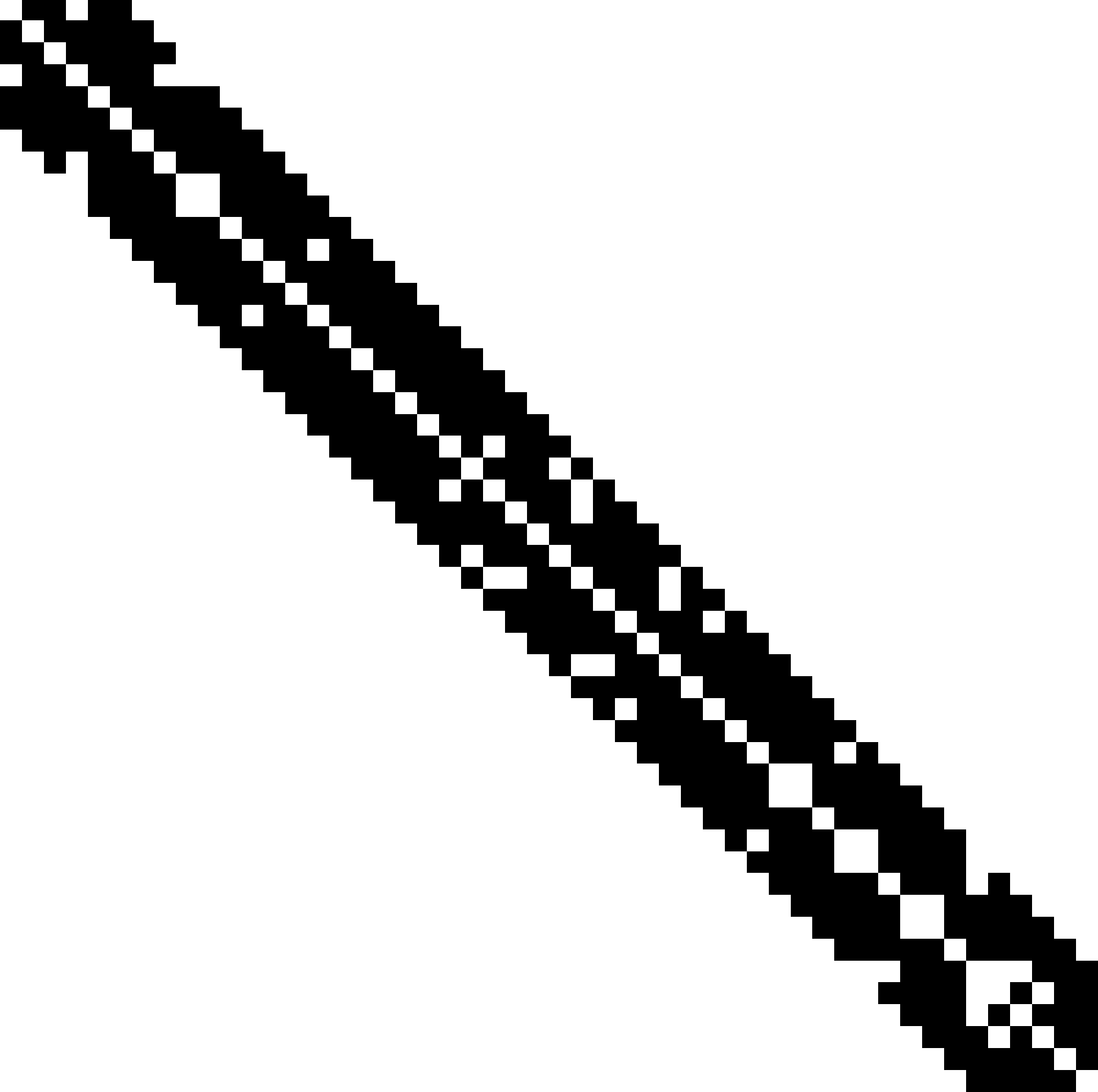}
\end{minipage}
\begin{minipage}{0.25\textwidth}
\includegraphics[width=1\linewidth]{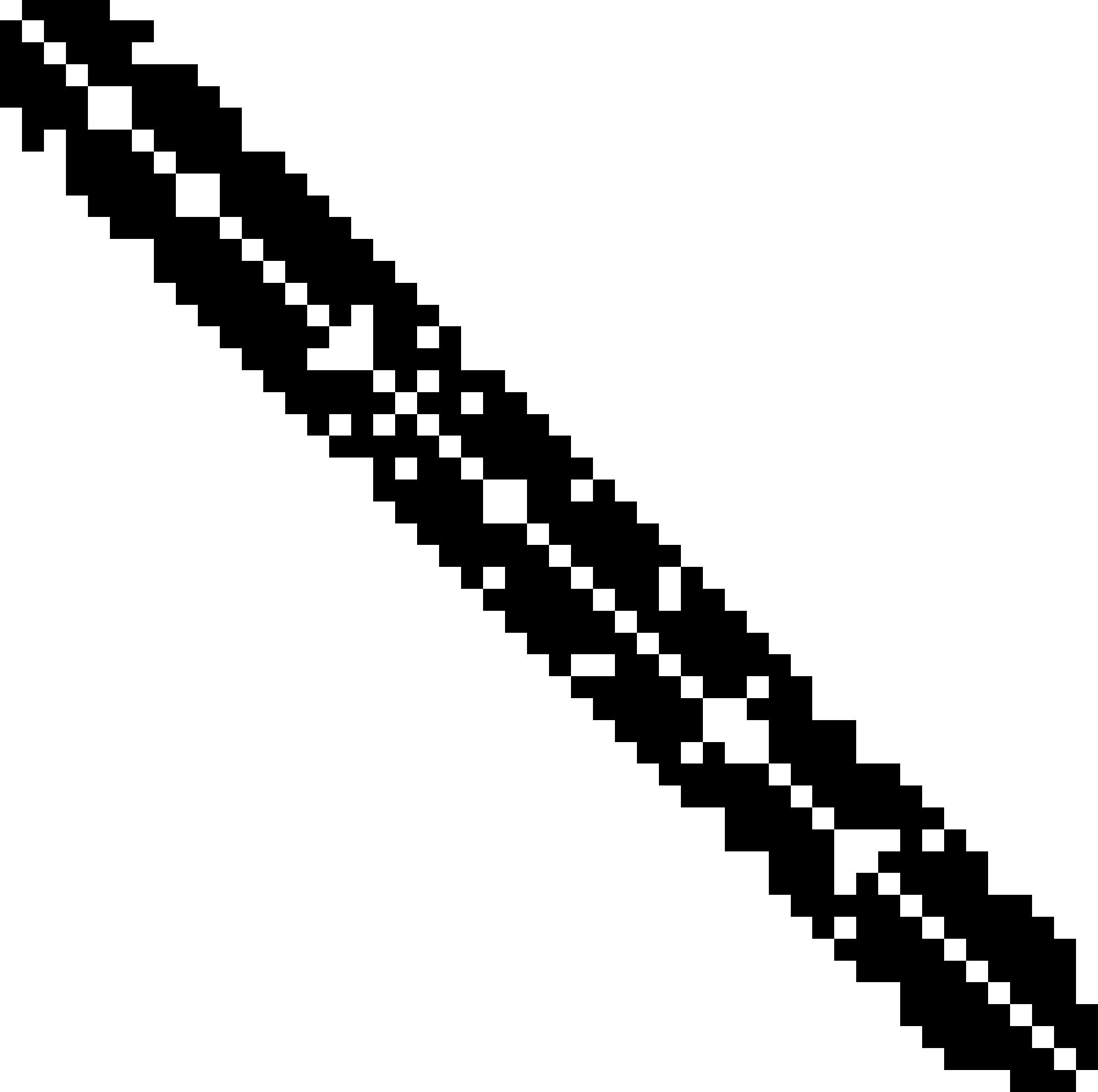}
\end{minipage}
\begin{minipage}{0.25\textwidth}
\includegraphics[width=1\linewidth]{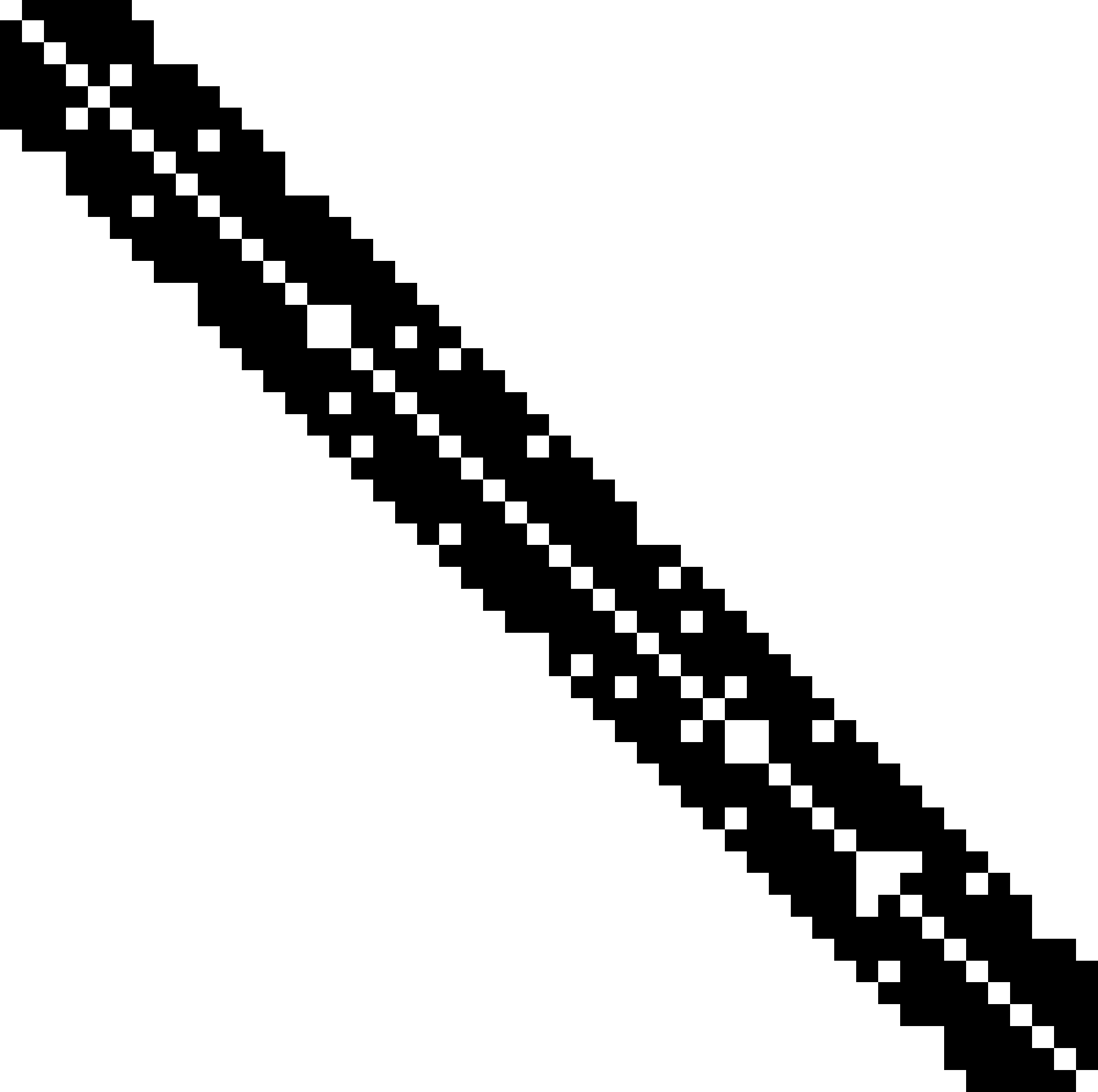}
\end{minipage}\\
\textcolor{magenta}{3-hub: }
\begin{minipage}{0.25\textwidth}
\includegraphics[width=1\linewidth]{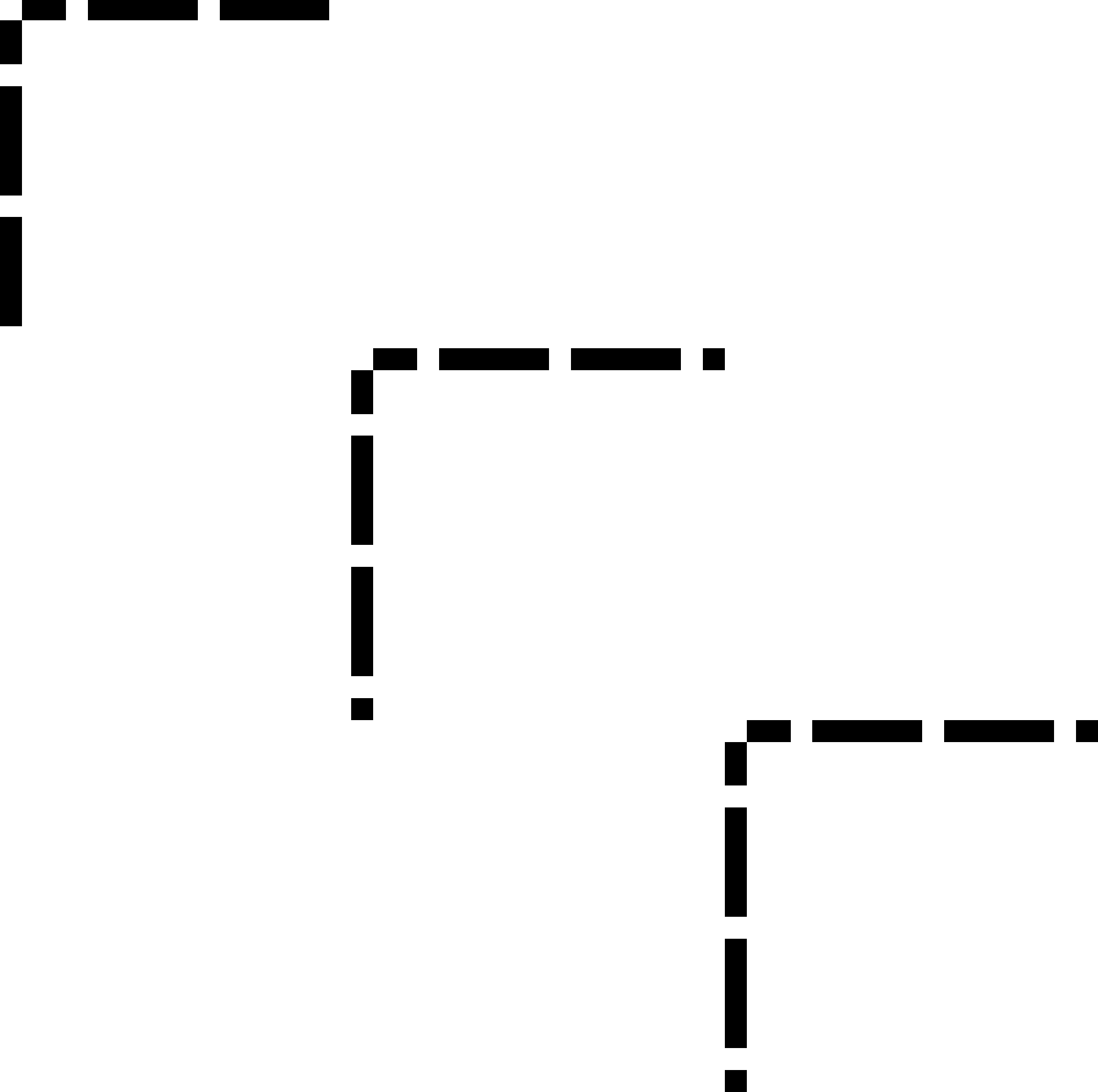}
\end{minipage}
\begin{minipage}{0.25\textwidth}
\includegraphics[width=1\linewidth]{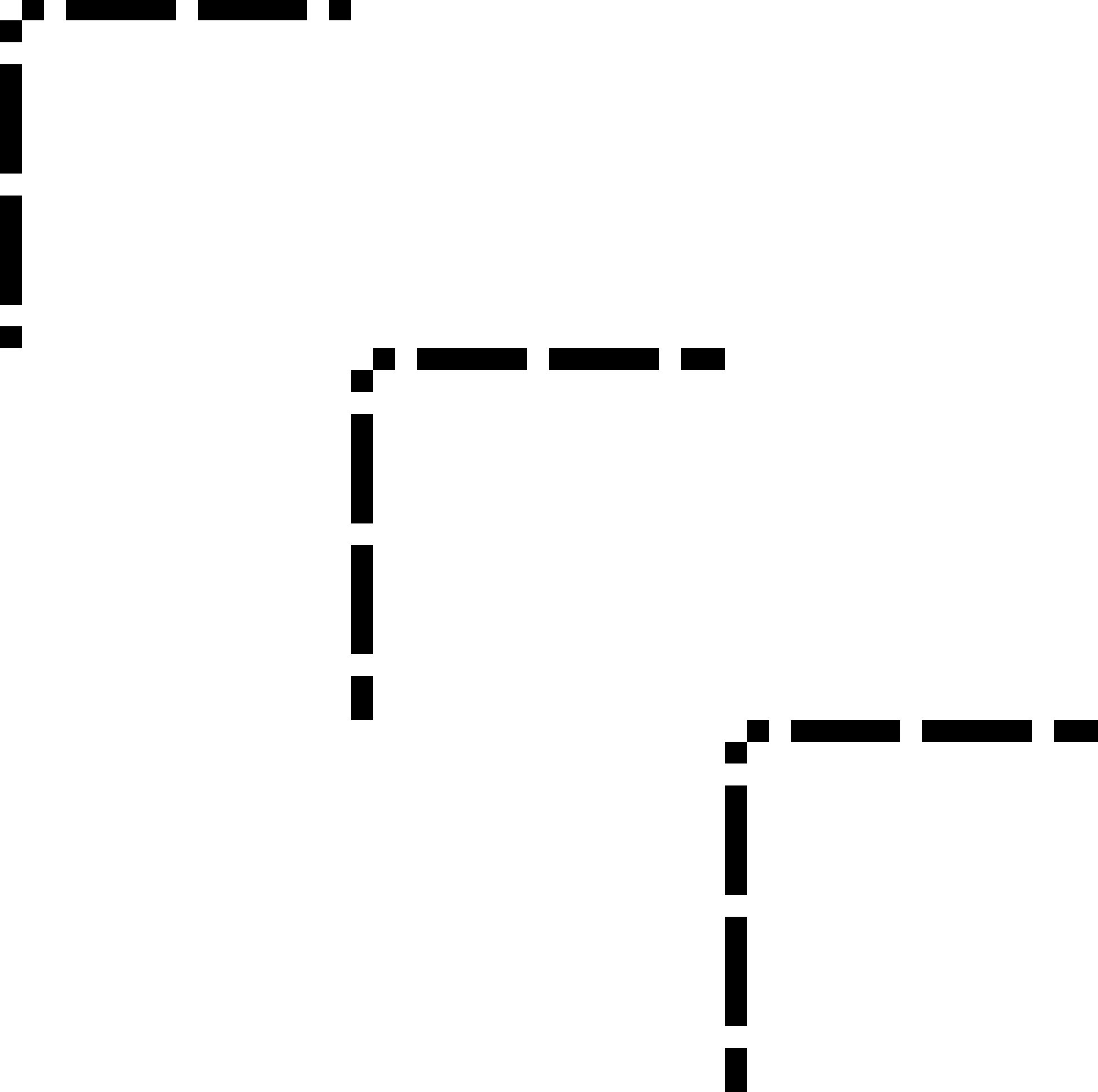}
\end{minipage}
\begin{minipage}{0.25\textwidth}
\includegraphics[width=1\linewidth]{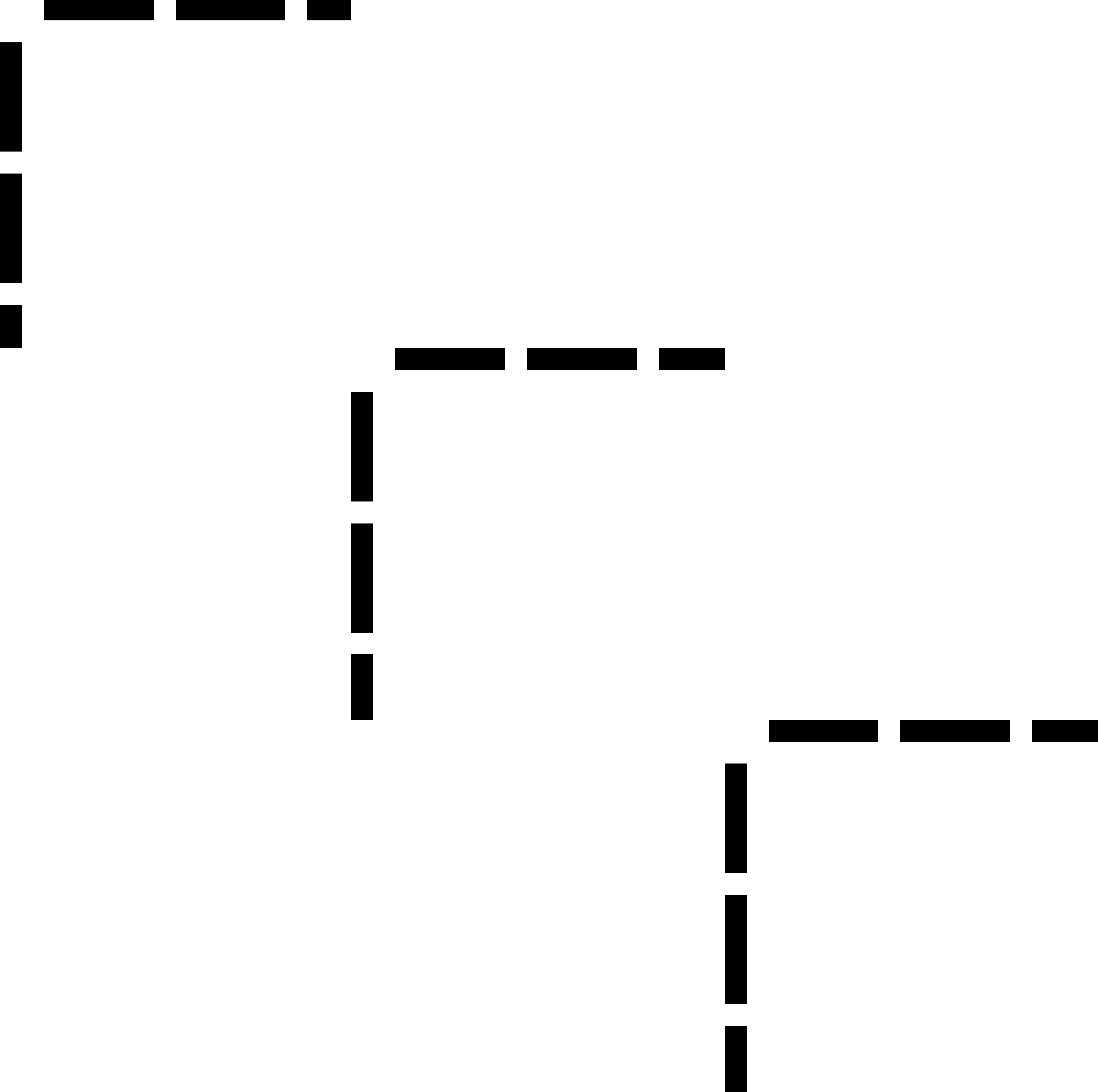}
\end{minipage}
\caption{Adjacency matrices of the 3 GGMs to be compared 
}\label{fig:sim.adj}
\end{center}
\end{figure}

The simulation schemes are summarized in Table \ref{tab:sim2}. For simplicity, we kept $n$ the same across the 3 graphs.   For each graph type, we first applied R function \texttt{simGraph} to simulate two $p\times p$ matrices $A_0$ and $B^{(l)}$ ($l=1,2,3$), where $A_0$ acts as the ``baseline'' structure and  $B^{(l)}$ characterizes the ``deviation'' from the baseline in graph $l$. The adjacency matrix  $A^{(l)}$ was constructed as $A^{(l)}= A_0+1(A_0\ne B^{(l)})$, where $1()$ is the indicator function. 
The adjacency matrices $A^{(l)}$ simulated for the GGM  case are shown in Figure \ref{fig:sim.adj}. Each graph is relatively sparse, either in the banded- and 3-hub graphs; and the differences in edges across are sparse as well; in other words, the 3 graphs look similar expect for a small proportions of different edges (the last 3 columns in Table \ref{tab:sim2}).
Given $A^{(l)}$, we then simulated the nodes values Gaussian and Poisson $\X$, respectively, using R function \texttt{XMRF.Sim}.
\begin{table}[!htb]
\begin{center}
\begin{tabular}{llll ccc l ccc}
\hline
graph & n & p &total&\multicolumn{3}{c}{non-zero edges}&& \multicolumn{3}{c}{different edge patterns in}\\
type &  &  &   edges  &\multicolumn{3}{c}{} &&\multicolumn{3}{c}{each pair of graphs}\\
\cline{5-7} \cline{9-11}
& & &  & G1& G2& G3 && G1:G2 & G1:G3 &G2:G3 \\
\hline
GGM (scale-free)& 100 & 50 &1225 & 224&224&225  && 66&65&69\\
GGM (banded)& 100 & 50 &1225 & 214&208&217  && 38&35&39\\
GGM (hub)& 100 & 50 &1225 & 38&38&38  && 18&18&18\\
PGM (scale-free)& 50  & 50 &1225 & 297&280&276  && 59&68&59\\
\hline
\end{tabular}
\caption{Simulation schemes for multiple graph models} \label{tab:sim2}
\end{center}
\end{table}

We run 100 repetitions in each graph-type case. For PANDAm-NS, we used the lasso-type noise  $e_1$ to yield edge sparsity in each graph and attempted both JGL and JFR-type noises $e_2$ to promote similarity across graphs. For the GGM case, we also employed the SCIO-based approach to construct the graphs in addition the NS approach. The specifications of the tuning and algorithm parameters in PANDAm are given in Table \ref{tab:sim2par}. $\lambda_1$, the tuning parameter associated with the lasso-type noise $e_1$ is left to vary so to examine how the regularization effect and the estimation accuracy change with  $\lambda_1$ and $\lambda_2$, which is 1/4 of $\lambda_1$, in PANDAm. As $n_{e_1}$ and $n_{e_2}$ are fixed, larger $\lambda_1$ and $\lambda_2$ imply more sparsity in each graph and in dissimilarity across the graphs.
\begin{table}[!htb]
\begin{center}
\resizebox{\columnwidth}{!}{
\begin{tabular}{lcccc ccccc ccc}
\hline
graph (method) & $\gamma$ & $\sigma^2$ &$\lambda_2$ & $T$ &  $n_{e,1}$(JGL)& $n_{e,2}$(JGL)&$n_{e,1}$(JFR) &$n_{e,2}$(JFR) & $m$ & $\tau_0$\\
\hline
GGM (NS)& 1 & 0 &$\lambda_1/4$ & 80 & 4,000&4,000 &2,000&2,000 & 1 & $10^{-4}$\\
GGM (SCIO)& 1 & 0 &$\lambda_1/4$ & 150 & 2,000&2,000 &2,000&2,000 & 1 & $10^{-4}$\\
PGM (NS)& 1 & 0 &$\lambda_1/4$ & 80 & 4,000&4,000 & 2,000&2,000 & 2 & $10^{-4}$\\
\hline
\end{tabular}}
\caption{Tuning and algorithm parameters in PANDAm-NS in the simulation study}\label{tab:sim2par}
\end{center}
\end{table}

\begin{figure}[!htb]
\begin{center}
\begin{minipage}{0.4\textwidth}
\includegraphics[width=1\linewidth]{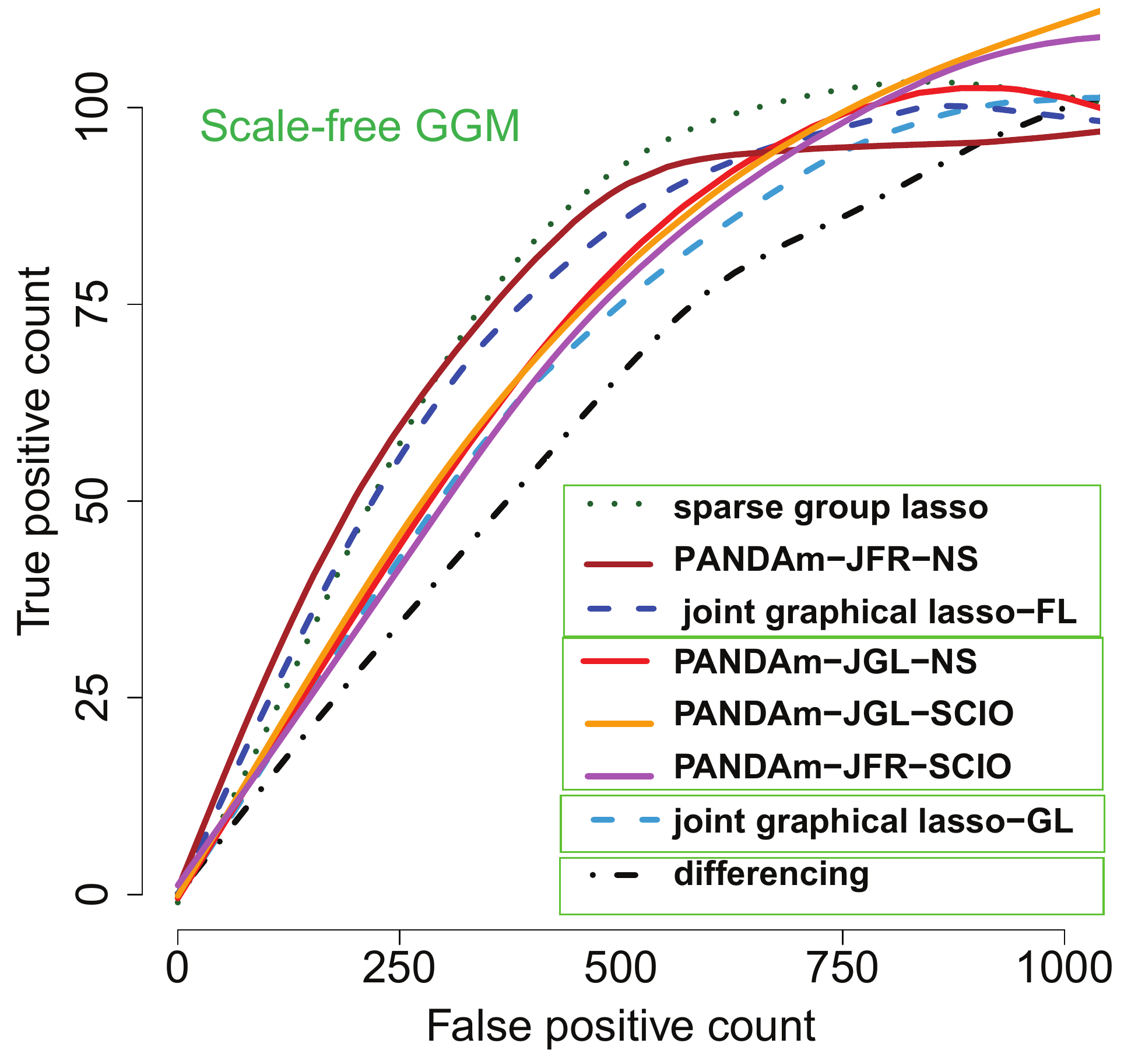}
\end{minipage}
\begin{minipage}{0.4\textwidth}
\includegraphics[width=1\linewidth]{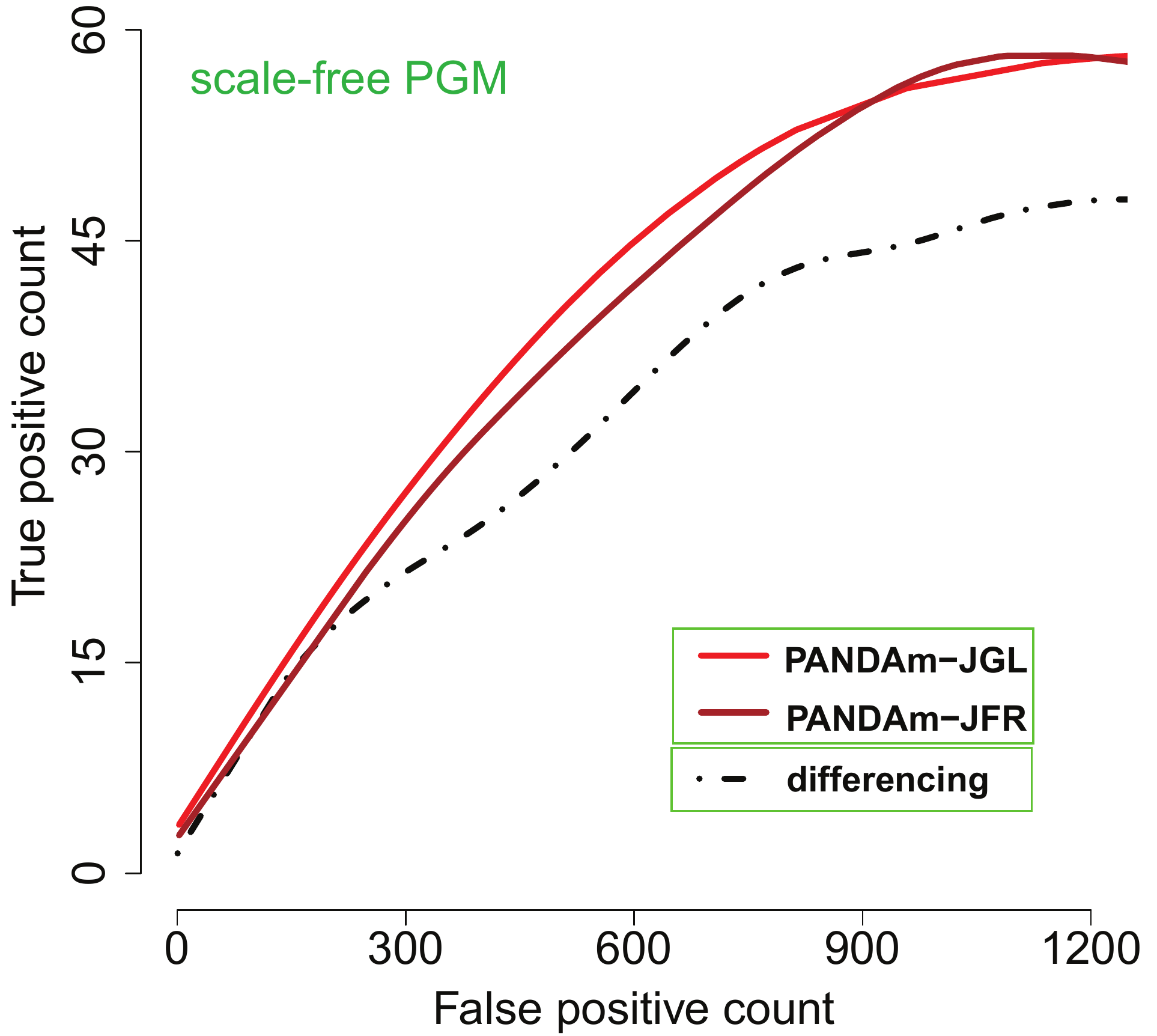}
\end{minipage}
\begin{minipage}{0.4\textwidth}
\includegraphics[width=1\linewidth]{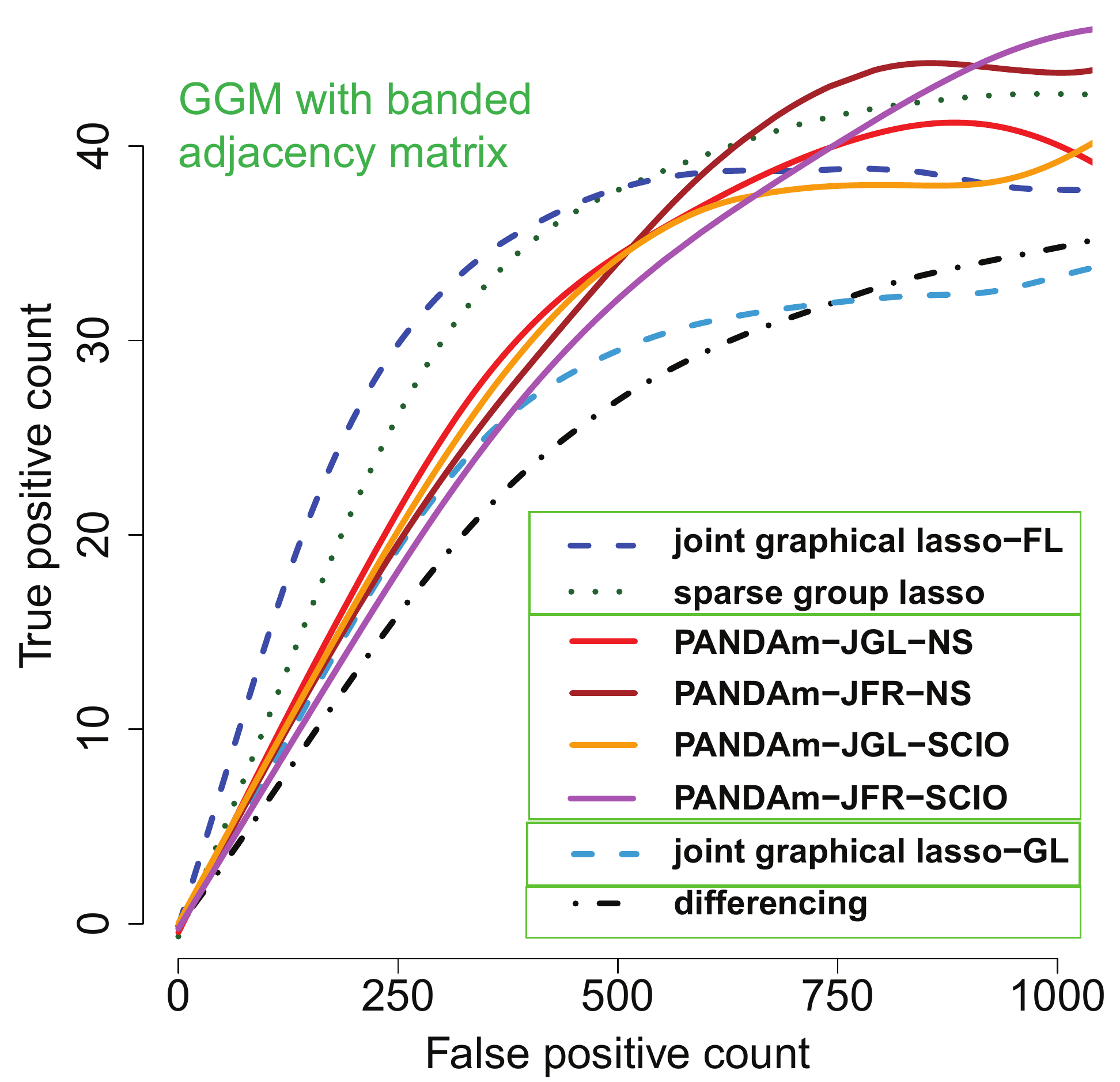}
\end{minipage}
\begin{minipage}{0.4\textwidth}
\includegraphics[width=1\linewidth]{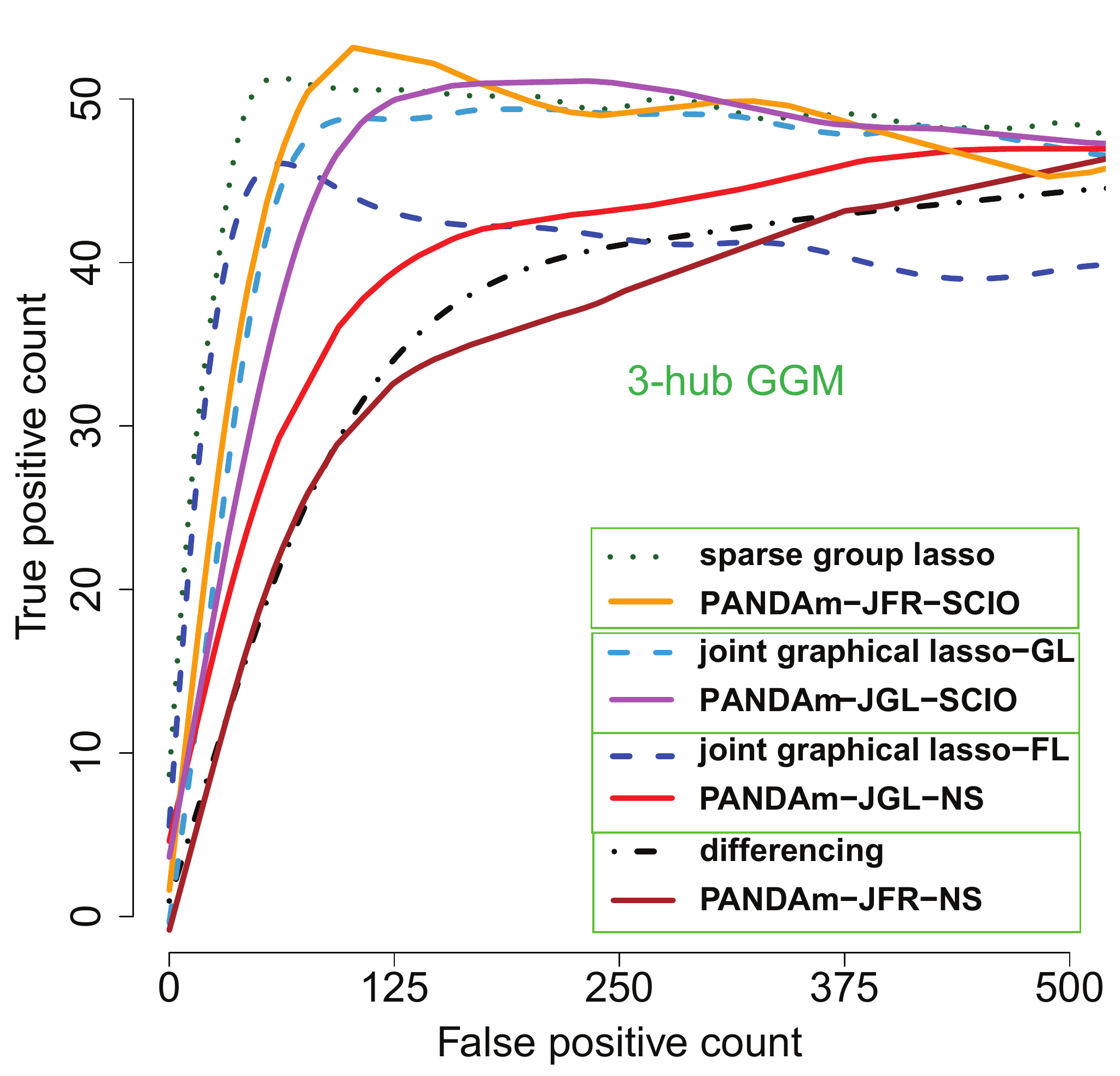}
\end{minipage}
\caption{ROC curves in the simulation study on multiple graph construction. The approaches in the legend of each plot are arranged in a descending order in performance; and boxed  boxed together if they have similar performance.}\label{fig:ROC}
\end{center}
\end{figure}

Figure \ref{fig:ROC} depicts the ROC curves with the false positive counts (FP) and true positive counts (TP) for different approaches. ``Positive'' is defined as detecting a disagreement in edge connection between two nodes in between two graphs. The FP and TP shown in the figures are the sum of FP and the sum of TP, respectively, over the 3 pairwise comparisons among the 3 graphs.  For PANDAm, we examine  a grid of $\lambda_1$ values to generate a range of FP and TP for the ROC curves. For the joint graphical lasso and the SGL approaches, we also vary the tuning parameter $\lambda_1$ that controls the sparsity of each graph, and set the tuning parameter for controlling the similarity across  the graphs at 1/4$\lambda_1$. For the na\"ive differencing, since each graph is estimated separately, there is only one tuning parameter, which controls the sparsity of each graph and is set to the same across the 3 graphs for each graph type.

\emph{First}, as expected, the na\"ive differencing approach performs the worst across all cases, with  lower TP for a given FP compared to the joint training methods (PANDAm, joint graphical lasso, and sparse group lasso).
\emph{Second}, in the construction of the scale-free GGMs, PANDAm-JFR-NS, NS with the sparse group lasso penalty, and the joint graphical lasso with the fused lasso penalty seem to perform the best for small FP; followed by PANDAm-NS-JGL, PANDAm-SCIO-JGL and PANDAm-SCIO-JFR, which delivered very similar performance. The joint graphical lasso  with the group lasso penalty seems to be slightly worse compared to the above 3 PANDAm approach in this simulation setting, which tend to surpass the joint graphical lasso approach with the group lasso penalty as FP increases.
\emph{Third}, in the construction of the GGMs with the banded adjacency matrices, the joint graphical lasso with the fused lasso penalty and the sparse group lasso seem to be the best at the small FP range, followed by all the PANDAm techniques and the joint graphical lasso approach with the group lasso penalty.
\emph{Fourth}, in the construction of the GGMs with the 3 hub nodes, the sparse group lasso and PANDAm-JFR-SCIO lead the performance; the joint graphical lasso with the group lasso penalty, and PANDAm-JGL-SCIO follow with a similar performance. PANDAm-JGL-NS and PANDAm-JFR-NS preformed slightly better than the differencing approach.
\emph{All taken together}, the sparse group lasso adapted for estimating graph models is the best for simultaneously constructing GGMs for the examined network structures. The performance between PANDAm and the joint graphical lasso alternates, depending on the network structure and the specific regularizer employed by PANDAm and the joint graphical lasso, respectively. Among all the joint estimation approaches examined in this study, only the joint graphical lasso  leads to positive definite estimates for the precision matrix in the GGM case.

\section{Case Study: the Lung Cancer Microarray Data}\label{sec:case}
To demonstrate the use of the PANDAm technique in the estimation and differentiation of multiple UGMs, we apply PANDA to a real-life data set with protein expression levels of subjects diagnosed with acute myeloid leukemia (AML). 
The data set is available from the \href{http://bioinformatics.mdanderson.org/Supplements/Kornblau-AML-RPPA/aml-rppa.xls}{MD Anderson Department of Bioinformatics and Computational Biology}. The subjects in the data are classified by AML subtype according to the French-American-British (FAB) classification system based on the criteria including cytogenetics and cellular morphology. We focus on 4 AML subtypes out of the 11 subtypes recorded in the data, M0 (17 subjects), M1 (34 subjects), M2 (68 subjects), and M4 (59 subjects). Data on the other 7 AML subtypes are not used because of the small sample sizes of these subtypes ($5 \sim 13$).  

The similarity across the protein networks from difference AML subtypes are often of interest, as the knowledge about common protein interactions across various AML subtypes can be helpful for developing treatments for AML. PANDAm is used to jointly construct 4 GGMs for the 4 AML subtypes (M0, M1, M2, M4) among 18 proteins (nodes) that are known to be involved in the apoptosis and cell cycle regulation according to the KEGG database \citep{kanehisa2011kegg}. The protein expression levels of the 18 proteins are assumed to follow Gaussian distributions. Because the main objective is to find common edges  across the 4 GGMs, the JGL regularization was used to promote similar non-zero patterns. We set both $n_{1,e}$ and $n_{2,e}$ at 600. For the tuning parameters, we specified $\gamma=1$ and $\sigma=0$ to obtain a lasso-type penalty and selected $\lambda_1^{(1)}=0.005$, $\lambda_1^{(2)}=0.010$, $\lambda_1^{(3)}=0.022$, $\lambda_1^{(4)}=0.022$ and $\lambda_2=0.018$ using the  extended BIC criterion \citep{chen2008extended, foygel2010extended, haslbeck2015structure} with a grid search. We ran 50 iterations in R (version 3.4.0) on the Linux x86\_64 operating system. The computation took approximately 46 seconds. 

\begin{figure}[!htb]
\begin{center}
\includegraphics[width=0.85\textwidth]{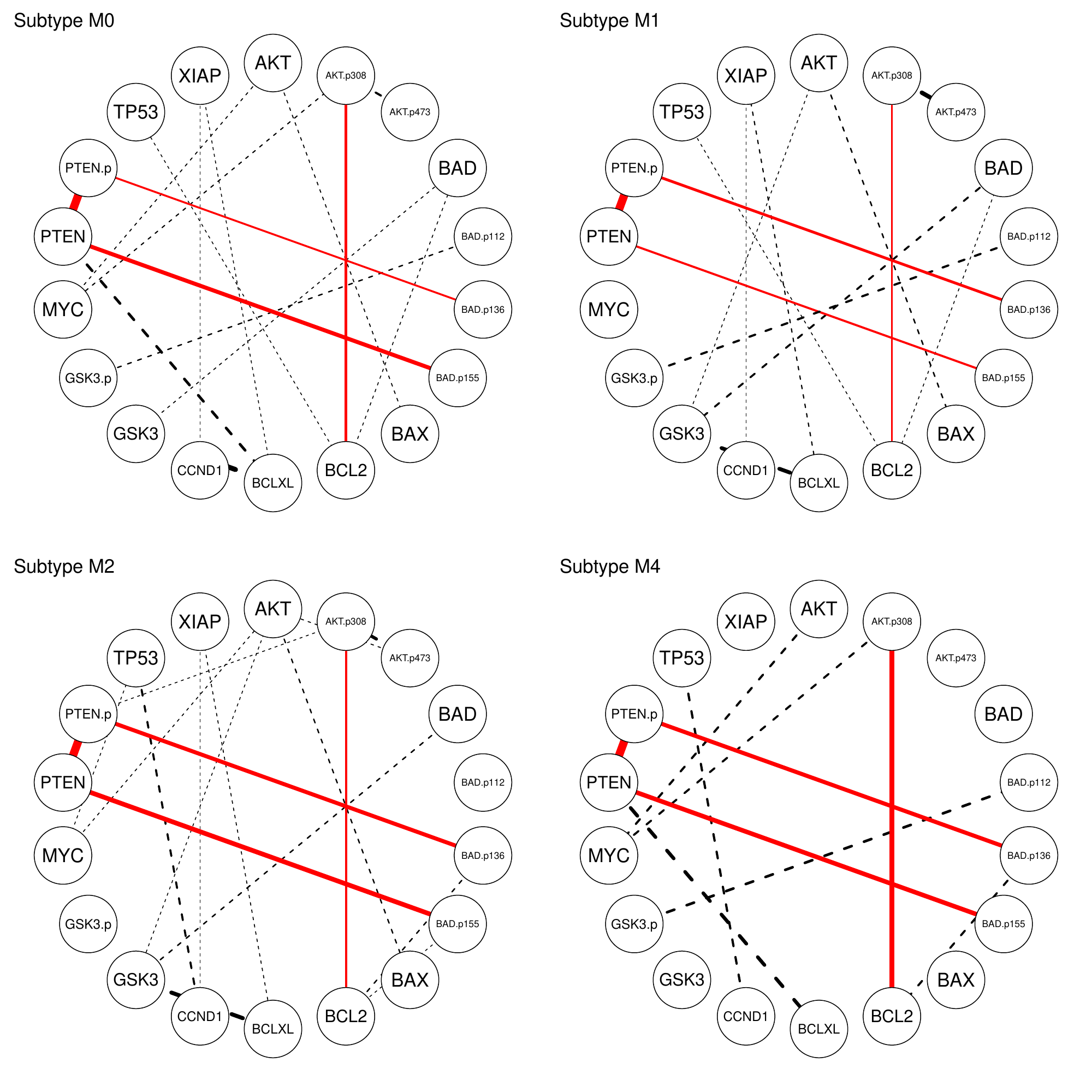}
\caption{\label{fig:qgraph_4ggm}
Joint estimation via PANDAm-NS-JGL of GGMs in four subtypes of AML (the edge width is proportional to its weight; the red edges are common to all 4 UGMs while the dashed black edges are not).}
\end{center}
\end{figure}

Figure \ref{fig:qgraph_4ggm} displays the four GGMs estimated using PANDAm-NS-JGL. The red edges indicate protein interactions common to all 4 AML subtypes (M0, M1, M2 and M4), and
four are identified, including the edges between ``PTEN'' and  ``PTEN.p'', ``PTEN'' and  ``BAD.p155'', ``PTEN.p'' and ``BAD.p136'', and ``AKT.p308'' and ``BCL2''. The shared structure we obtained using PANDAm was also identified in previous studies on protein networks of AML subtypes \citep{peterson2015bayesian}.
In addition, for the  proteins involved in the shared structure, proteins ``PTEN'' and  ``PTEN.p'' had been shown to have relatively high expression in the four AML subtypes we consider here, compared with other subtypes such as M6 and M7 \citep{kornblau2009functional}. In subtypes M0, M1, and M2, the expression levels of proteins ``BAD.p136'' and ``BCL2'' tend to be high, and that of ``AKT.p308'' tends to be low. The shared structure in the estimated graphs could shed light on similarities in the structures of the protein networks across AML different subtypes, which in turn will guide treatment development for AML.

\section{Discussion} \label{sec:discussion}
We extend the data augmentation regularization technique PANDA for constructing a single graph to jointly estimating multiple graphs, referred to as PANDAm. Toward that end, we design two types of noises to augment the observed data from multiple graphs. The first type regularizes the structure of each graph such as with sparsity while the second type promotes either the structural similarity (the JGL regularization) or the numerical similarity (the JFR regularization) among the edges in the same positions across the graphs.

To the best our knowledge, our extension of PANDA to the multiple graph setting offers the first approach for jointly constructing multiple UGMs in general (GGMs and MGMs included), when the conditional distribution of each node in each graph given the other nodes can be modelled by an exponential family. For general UGMs, we implement PANDAm in the NS framework to estimate the model parameters; for GGMs, there are other alternatives to the NS framework, such as through the Cholesky decomposition of the precision matrix or through the SCIO estimator. In all cases, we establish the expected regularization effect in PANDAm.  Computationally, PANDAm  runs GLMs to obtain MLE iteratively on the combined augmented  data across graphs, and the algorithms can be programmed with a few lines in any software that offers built-in GLM functions. When the augmented data are large, the computation could take some time but usually the PANDA algorithms does not require many iterations to converge.

There are two penalty terms in PANDAm. Users need to choose at least two tuning parameters in the implement of PANDAm. First, users need to first to decide which NGDs to use for the first and the second types noise. For the first type,  we refer users  to \citet{panda1}. For the second type, it does not seem there is much difference between the JFL and JGL regularization in the simulation studies despite their conceptual disparity. Once the NGDs are determined, the collection of all tuning parameters can be chosen by cross validation or information-based criteria (see \citet{KL}) via a grid search, such as the extended BIC as we employed in the simulation studies, to choose the tuning parameters lead to the better identification of the graphical structures or to better predictive power.  On the other hand, \citet{Danaher2014} states there is no `` theoretically optimal value'' for the parameters that controls the relative separate vs. simultaneous regularization across graphs since ``the optimal value would need to be a function of the number of covariates and group sizes among other things''.  

PANDAm enjoys the theoretical properties established by \citet{panda1} in the single graph setting,  including the Gaussian tail bound of the augmented loss function, and the almost sure convergence of the noise-augmented loss function to its expectation, which is a penalized loss function with the designed regularization effect. On the other hand, we do investigate whether the estimated graphs are structurally consistent for true graph structures, which this can be a future research topic, though a likely challenging one.   To the best of our knowledge, there is no theoretical work on that topic yet, not even for GGMs, which have nice connection with the precision matrices of Gaussian distributions. For example, \citet{Danaher2014}, while proposing the joint graphical lasso for jointly estimating multiple GGMs, do not investigate establish the structural consistency of the jointly  estimated GGMs for true structures. One possible direction is to borrow the work by   \citet{yang2014} and \citet{UGMEXP2015}  in the single graph setting, which connect the  conditional distribution of each node given others with UGMs via exponential families and GLMs, and shows the graph constructed by minimizing the regularized loss function,  with  $l_1$ penalty terms placed on the regression coefficients, is structurally consistent for the underlying UGM.

\bibliographystyle{apa}

\section*{Appendix}
\appendix
\numberwithin{equation}{section}
\setcounter{equation}{0}

\section{Proof of Proposition \ref{prop:ElossUGM}}\label{app:prop1}
The proof  is similar to the proof for Proposition 2 in \citet{panda1} on the regularization effect of PANDA for a single UGM. The only difference in PANDAm is the additional penalty  $P_2(\Theta)$  brought by the second type of noise $\e_2$. WLOG, we derive $P_2(\Theta)$ in the framework of multiple GGMs. The proof can be easily combined with the proof for Proposition 2 in \citet{panda1} to obtain $P_2(\Theta)$ for UGMs in general.
\begin{align}
&\E_{\e_1,\e_2}\left(l_p(\Theta|\x,\e_1,\e_2)\right)\!=\!\sum_{l=1}^{q}\!\sum_{i=1}^{n_l}\!\sum_{j=1}^{p}\!\left(\!x_{ij}^{(l)}\!-\!\!\sum_{k\ne j}x_{ik}^{(l)}\theta^{(l)}_{jk}\!\right)^{\!2}\!\!\!+\!\E_{\e_1,\e_2}\sum_{h=1}^{2}\!
\sum_{i=1}^{n_{e,h}}\!\sum_{j=1}^{p}\!\left(\!e_{ijj,h}\!-\!\!\sum_{l=1}^{q}\!
\sum_{k\ne j}e_{ijk,h}^{(l)}\theta_{jk}^{(l)}\!\right)^{\!2}\notag\\
&\textstyle=l(\Theta|\x)+\!
n_{e,1}\!\sum_{j=1}^{p}\!\sum_{l=1}^{q}\!
\sum_{k\ne j}\E_\e\left(e_{ijk,1}^{(l)2}\right)\theta_{jk}^{(l)2}+\notag\\
&\quad\boxed{\textstyle n_{e,2}\!\sum_{j=1}^{p}\!\sum_{k\ne j}\left\{\sum_{l=1}^{q}\!\E_\e\left(e_{ijk,2}^{(l)2}\right)\theta_{jk}^{(l)2}+\sum_{l=1}^{q}\!\sum_{v=1}^{l-1}2\E\left(e_{ijk,2}^{(l)}e_{ijk,2}^{(v)}\right)\theta_{jk}^{(l)}\theta_{jk}^{(v)}\right\}}.\label{eqn:boxed}
\end{align}
There are no cross-product terms among $e_1$ because all $e_1$ noise terms are independent. When $e_2$ is of the JGL type, all $e_2$ terms are also independent, so the cross-product terms among $e_2$ are 0 and the boxed term above becomes
\begin{align*}
&\textstyle n_{e,2}\!\sum_{j=1}^{p}\!\sum_{k\ne j}\sum_{l=1}^{q}\!\E_\e\left(e_{ijk,2}^{(l)2}\right)\theta_{jk}^{(l)2}=
n_{e,2}\!\sum_{j=1}^{p}\!\sum_{k\ne j}\sum_{l=1}^{q}\!\V_\e\left(e_{ijk,2}^{(l)}\right)\theta_{jk}^{(l)2}\\
=&\textstyle n_{e,2}\lambda_2\!\sum_{j=1}^{p}\sum_{k\ne j}\sum_{l=1}^{q} \left(\sum_{l=1}^{q}\theta_{jk}^{(l)2}\right)^{-1/2}\theta_{jk}^{(l)2}=
n_{e,2}\lambda_2\sum_{j=1}^{p}\sum_{k\ne j}\left(\sum_{l=1}^{q}\theta_{jk}^{(l)2}\right)^{1/2}
\end{align*}
When $e_2$ is of the JFR type, the boxed terms in Eqn (\ref{eqn:boxed}) becomes
\begin{align}\label{eqn:jfr}
&\textstyle n_{e,2}\!\sum_{j=1}^{p}\!\sum_{k\ne j}\left(\sum_{l=1}^{q}\!\E_\e\left(e_{ijk,2}^{(l)2}\right)\theta_{jk}^{(l)2}+\sum_{l=1}^{q}\!\sum_{v=l-1}2\E\left(e_{ijk,2}^{(l)}e_{ijk,2}^{(v)}\right)\theta_{jk}^{(l)}\theta_{jk}^{(v)}\right).
\end{align}
The covariance matrix of $e_{ijk,2}^{(l)}$ for $l=1,\ldots,q$ is $\lambda_2 \mathbf{TT}'$, the diagonal element of which is $2\lambda_2$, and the off-diagonal elements lead the following covariance: $\mbox{cov}\left(e_{ijk,2}^{(1)}, e_{ijk,2}^{(2)}\right)=-2\lambda_2$ when $q=2$; $\mbox{cov}\left(e_{ijk,2}^{(l)}, e_{ijk,2}^{(l-1)}\right)=\mbox{cov}\left(e_{ijk,2}^{(q)}, e_{ijk,2}^{(1)}\right)=-\lambda_2$ for $l\ge2$ when $q\ge3$, and 0 otherwise. Plugging these terms into Eqn (\ref{eqn:jfr}), we have
\begin{align*}
&\textstyle n_{e,2}\!\sum_{j=1}^{p}\!\sum_{k\ne j}\left(2\lambda_2\theta_{jk}^{(l)2}+4\lambda_2\theta_{jk}^{(1)}\theta_{jk}^{(2)}\right)=
 2n_{e,2}\lambda_2\!\sum_{j=1}^{p}\!\sum_{k\ne j}\left(\theta_{jk}^{(1)}-\theta_{jk}^{(2)}\right)^2\mbox{ for } q=2\\
=&\textstyle n_{e,2}\lambda_2\!\sum_{j=1}^{p}\!\sum_{k\ne j}\left(\theta_{jk}^{(1)}-\theta_{jk}^{(2)}\right)^2\mbox{ (refine $\lambda_2\leftarrow 2\lambda_2$ as $\lambda_2$ itself is a tuning parameter)}\\
&\textstyle n_{e,2}\!\sum_{j=1}^{p}\!\sum_{k\ne j}\left(2\lambda_2\sum_{l=1}^q\theta_{jk}^{(l)2}+2\lambda_2\sum_{l=2}^q\theta_{jk}^{(l)}\theta_{jk}^{(l-1)}+2\lambda_2\theta_{jk}^{(q)}\theta_{jk}^{(1)}\right)\qquad\qquad \mbox{for } q\ge3\\
=&\textstyle n_{e,2}\lambda_2\!\sum_{j=1}^{p}\!\sum_{k\ne j}
\sum_{l,v\in\mathcal{S}}\left(\theta_{jk}^{(l)}\!-\!\theta_{jk}^{(v)}\right)^{\!2}
\end{align*}
Denote by $\mathcal{S}$  the combinatorics set $(_2^q)$ among the $q$ graphs. All taken together, when $e_2$ is of the JFR type, the boxed term in Eqn (\ref{eqn:boxed}) is
$$ \textstyle \lambda_2 n_e \sum_{j=1}^{p}\sum_{k\ne j}\sum_{l,v\in\mathcal{S}} \!\left(\theta_{jk}^{(l)}\!-\!\theta_{jk}^{(v)}\right)^{\!2} $$

\section{Proof of Proposition \ref{prop:SCIO}}\label{app:mcolumnwiseSCIO}
\vspace{-12pt}
\begin{align*}
&\textstyle\E_\e(l_p(\Theta|\x,\e_1,\e_2))=(2n)^{-1}\sum_{j=1}^{p}\Theta_j\E_\e\left(\tilde{\x}_j^{'}\tilde{\x}_j\right)\Theta_j-\sum_{j=1}^{p}\Theta_j\Xi_j\\
=&\textstyle\sum_{j=1}^{p}\Theta_j\left(\frac{1}{2n}{\x}_j^{'}{\x}_j+\E_\e\left(\e_{j,1}^{'}\e_{j,1}\right)+\E_\e\left(\e_{j,2}^{'}\e_{j,2}\right)\right)\Theta_j-\sum_{j=1}^{p}\Theta_j\Xi_j\\
=&\frac{1}{2n}\!\sum_{j=1}^{p}\Theta_j({\x}_j^{'}{\x}_j)\Theta_j\!-\!\sum_{j=1}^{p}\Theta_j\Xi_j\!+\!\sum_{j=1}^{p}\!\sum_{k\neq j}\!\sum_{h=1}^{2}\!n_{e,h}\!\sum_{l=1}^{q}\!\sum_{v=\{l-1,l,l+1\}}\!\!\!\mbox{Cov}\left(e_{ijk,h}^{(l)}e_{ijk,h}^{(v)}\right)\theta_{jk}^{(l)}\theta_{jk}^{(v)}\\
=&l(\Theta|\x)\!+\!\lambda_1 n_{e,1}\!\!\sum_{j=1}^{p}\!
\sum_{k\ne j}\!\sum_{l=1}^{q}\left|\theta_{jk}^{(l)}\right|^{2-\gamma}+\begin{cases}
\lambda_2n_{e,2}\sum_{j=1}^{p}
\!\sum_{k\ne j}\left(\sum_{l=1}^{q}\theta_{jk}^{(l)2}\right)^{1/2} & \mbox{for JGL}\\
\lambda_2n_{e,2}\sum_{j=1}^{p}\!\sum_{k\ne j}\sum_{l,v\in\mathcal{S}} \!\left(\theta_{jk}^{(l)}\!-\!\theta_{jk}^{(v)}\right)^{\!2} & \mbox{for JFR}
\end{cases}
\end{align*}
where $\mathcal{S}$ denotes the combinatorics set $(_2^q)$ among the $q$ graphs.

\clearpage
\setcounter{section}{0}
\setcounter{algorithm}{0}
\renewcommand\thesection{S.\arabic{section}}
\renewcommand\thealgorithm{S.\arabic{algorithm}}
\setstretch{2}
\begin{center}

\Large  \textbf{Supplementary Materials to }\\
\textbf{\emph{AdaPtive Noisy Data Augmentation (PANDA) for Simultaneous Construction Multiple Graph Models}}\\
\normalsize Yinan Li$^{1}$, Xiao Liu$^{2}$, and Fang Liu$^1$\\
\small$^1$ Department of Applied and Computational Mathematics and Statistics\\
\small$^2$ Department of Psychology\\
\small University of Notre Dame, Notre Dame, IN 46556, U.S.A.
\end{center}
\clearpage

\setstretch{1}

\section{PANDAm-CD Algorithm for multiple GGM estimation}
\begin{algorithm}[!htb]
\caption{PANDAm-CD for joint estimation of $q$ GGMs}\label{alg:CDJGLJFR}
\begin{algorithmic}[1]
\State \textbf{Input}:
\begin{itemize}[leftmargin=0.18in]\setlength\itemsep{-1pt}
\item random initial estimates $\bar{\bs{\theta}}_j^{(l)(0)}$ for $j=1,\ldots,p$ and $l=1,\ldots,q$.
\item a NGD for generating $e_1$  (Eqn (\ref{eqn:e1})), a NGD for $e_2$ (Eqns(\ref{eqn:JGLUGMe2}) or (\ref{eqn:JFRUGMe2})), maximum iteration $T$, noisy data sizes $n_{e,1}$ and $n_{e,2}$, thresholds $\tau,\tau_0$, width of moving average (MA) window $m$, banked parameter estimates after convergence $r$, inner loop $K$ in alternatively estimating $\bs{\theta}_j$ and $\sigma_j^2$
\end{itemize}
\State $t\leftarrow 0$; convergence $\leftarrow 0$
\State \textbf{WHILE} $t<T$ \textbf{AND} convergence $= 0$
\State \textbf{\hspace{0.3cm} FOR} $j = 2$ to $p$
\State \textbf{\hspace{0.6cm} FOR} $k =1:K$
\begin{enumerate}[leftmargin=0.5in]\setlength\itemsep{-1pt}
\item[a)] Generate $e_{ijk,1}^{(l)}$  for $i=1,\ldots,n_{e,1}$, and $e_{ijk,2}^{(l)}$ for $i=1,\ldots,n_{e,2}$, with $\bar{\bs{\theta}}_j^{(l)(t-1)}$ plugged in the variance terms of the NGDs. Combine $x^{(l)}, \e_1^{(l)}$ and $\e_2^{(l)}$ for $l=1,\ldots,q$ to obtain the augmented data as given in Figure \ref{fig:pandaCD_JGL}.
\item[c)] Obtain the OLS estimates $\hat{\bs{\theta}}_{j}^{(t)}=\left(\hat{\bs{\theta}}_{j}^{(1)(t)},\ldots,\hat{\bs{\theta}}_{j}^{(q)(t)}\right)$ from the regression model
$\tilde{\x}_{j}=\tilde{\x}^{(1)}[,1:(j-1)]\bs\theta_{j}^{(1)}+\ldots+\tilde{\x}^{(q)}[,1:(j-1)]\bs\theta_{j}^{(q)}$, where $\tilde{\x}^{(l)}[,1:(j-1)]$ refers to the 1-st to the $(j-1)$-th columns of  $\tilde{\x}^{(l)}$ for $l=1,\ldots,q$
\item[d)] If $t>m$, calculate the MA $\bar{\bs{\theta}}^{(l)(t)}_j=m^{-1}\sum_{b=t-m+1}^t \hat{\bs{\theta}}_{j}^{(l)(b)}$; otherwise $\bar{\bs{\theta}}^{(l)(t)}_j=\hat{\bs{\theta}}^{(l)(t)}_j$ for $l=1,\ldots,q$. Calculate  $\hat{\sigma}_{j}^{2(l)(t)}$ via Eqn (\ref{eqn:CDsigma2}).
\end{enumerate}
\textbf{\hspace{0.6cm}End FOR}\\
\hspace{0.3cm} \textbf{End FOR}
\State \hspace{0.3cm} Calculate the overall loss function $\bar{l}^{(t)}=\sum_{j=1}^p\sum_{l=1}^qn^{(l)}\hat{\sigma}_{j}^{2(l)(t)}$.
\State \hspace{0.3cm} Apply a convergence criteria  to $\bar{l}^{(t)}$; Let convergence $\leftarrow 1 $ if the convergence is reached.
\State  \textbf{End WHILE}
\State Continue to execute the command lines 4 and 6  for another $r$ iterations, and record $\bar{\bs\theta}^{(l)(b)}_j$ for $b=t+1,\ldots,t+r$ and $l=1,\ldots,q$, calculate the degrees of freedom $\nu_j^{(t)}=\mbox{trace}(\x_j(\tilde{\x}'_j\tilde{\x}_j)^{-1}\x'_j)$ and $\hat{\sigma}_{j}^{2(b)}=$ SSE$^{(t)}_{j}/(n-\nu_j^{(b)})$. Let $\bar{\bs{\theta}}^{(l)}_{jk}=(\bar{\theta}_{jk}^{(l)(t+1)},\ldots,\bar{\theta}_{jk}^{(l)(t+r)})$.
\State Set $\hat{\theta}_{jk}^{(l)}=0$ if $\left(\big|\max\{\bar{\bs{\theta}}_{jk}^{(l)}\}\cdot\min\{\bar{\bs{\theta}}_{jk}^{(l)}\}\big|<\tau_0\right) \cap \left(\max\{\bar{\bs{\theta}}_{jk}^{(l)}\}\cdot\min\{\bar{\bs{\theta}}_{jk}^{(l)}\}\right)<0$  for $k> j$; otherwise, set $\hat{\theta}_{jk}=r^{-1}\sum_{b=t+1}^{t+r}\bar{{\theta}}_{jk}^{(b)(l)}$. Set $\hat{D}=r^{-1}\sum_{b=t+1}^{t+r}\diag(\hat{\sigma}_1^{2(b)},\ldots,\hat{\sigma}_p^{2(b)})$. Calculate $\hat\Omega^{(l)}=\!\hat{L}^{(l)'}\hat{D}\hat{L}^{(l)} $
\State \textbf{Output}: $\hat{\Omega}^{(l)}$.
\end{algorithmic}
\end{algorithm}

\newpage
\section{PANDAm-SCIO Algorithm for multiple GGM estimation}
\begin{algorithm}[!htb]
\caption{PANDAm-SCIO for joint estimation of $q$ GGMs}\label{alg:SCIO}
\begin{algorithmic}[1]
\State \textbf{Pre-processing}: standardize $\x^{(l)}$ for $1=1,\ldots,q$ separately.
\State \textbf{Input}:
\begin{itemize}[leftmargin=0.18in]\setlength\itemsep{-1pt}
\item random initial estimates $\bar{\bs{\theta}}_j^{(l)(0)}$ for $j=1,\ldots,p$ and $l=1,\ldots,q$.
\item a NGD to generate $e_1$ (Eqn (\ref{eqn:e1})), a NGD to generate $e_2$ (Eqns (\ref{eqn:e2JGL.SCIO}) or  (\ref{eqn:e2JFR.SCIO})), maximum iteration $T$, noisy data sizes $n_{e,1}$ and $n_{e,2}$, thresholds $\tau,\tau_0$, width of moving average (MA) window $m$, banked parameter estimates after convergence $r$.
\end{itemize}
\State $t\leftarrow 0$; convergence $\leftarrow 0$
\State \textbf{WHILE} $t<T$ \textbf{AND} convergence $= 0$
\State \textbf{\hspace{0.3cm} FOR} $j = 1:p$
\begin{enumerate}[leftmargin=0.5in]\setlength\itemsep{-1pt}
\item[a)] Generate $e_{ijk,1}^{(l)}$  for $i=1,\ldots,n_{e,1}$, and $e_{ijk,2}^{(l)}$ for $i=1,\ldots,n_{e,2}$, with $\bar{\bs{\theta}}_j^{(l)(t-1)}$ plugged in the variance terms of the NGDs.
\item[b)] Obtain augmented data $\tilde{\x}$ by combining the standardized $\x^{(l)}$ and $\sqrt{2n}\e^{(l)}_{j,1}$,$\sqrt{2n}\e^{(l)}_{j,2}$ from a) according to Figure \ref{fig:pandaJGL}.
\item[c)] Calculate $\hat{\Theta}_j^{(t)}=2n\left(\tilde{\x}_j^{'}\tilde{\x}_j\right)^{-1}\Xi_j$
and obtain $\hat{\Theta}_j^{(t)}=\left[\hat{\bs{\theta}}_{j}^{(1)(t)}//\ldots//\hat{\bs{\theta}}_{j}^{(q)(t)}\right]$.
\item[d)] If $t>m$, calculate the MA $\bar{\Theta}^{(1)(t)}_j=m^{-1}\sum_{b=t-m+1}^t \hat{\Theta}_{j}^{(1)(b)}$; otherwise $\bar{\Theta}^{(1)(t)}_j=\hat{\Theta}^{(1)(t)}_j$
\end{enumerate}
\hspace{0.45cm} \textbf{End FOR}
\State \hspace{0.45cm} Plug in $\bar{\Theta}^{(l)(t)}_j$ for $j=1,\ldots,p$ in the loss function in  Eqn (\ref{eqn:SCIOpelfJGL}) to obtain $\bar{l}^{(t)}$
\State \hspace{0.45cm} apply a convergence  criteria  to $\bar{l}^{(t)}$. Let convergence $\leftarrow 1$ if the convergence is reached.
\State  \textbf{End WHILE}
\State Continue to execute the command line 5  for another $r$ iterations, and record $\bar{\Theta}^{(l)(b)}_j$ for $b=t+1,\ldots,t+r$ and $l=1,\ldots,q$. Let $\bar{\bs{\theta}}^{(l)}_{jk}=(\bar{\theta}_{jk}^{(l)(t+1)},\ldots,\bar{\theta}_{jk}^{(l)(t+r)})$.
\State Set $\hat{\theta}^{(l)}_{jk}=\hat{\theta}_{kj}^{(l)}=0$
if $\left(\big|\max\{\bar{\bs\theta}^{(l)}_{jk}\}\cdot\min\{\bar{\bs\theta}^{(l)}_{jk}\}\big|<\tau_0\right) \cap \left(\max\{\bar{\bs\theta}^{(l)}_{jk}\}\cdot\min\{\bar{\bs\theta}^{(l)}_{jk}\}<0\right)$
or $\left(\big|\max\{\bar{\bs\theta}^{(l)}_{kj}\}\cdot\min\{\bar{\bs\theta}^{(l)}_{kj}\}\big|<\tau_0\right) \cap \left(\max\{\bar{\bs\theta}^{(l)}_{kj}\}\cdot\min\{\bar{\bs\theta}^{(l)}_{kj}\}<0\right)$  for $k\ne j$; otherwise, set $\hat{\theta}_{jk}=\min\{\bar{\bs\theta}^{(l)}_{jk},\bar{\bs\theta}^{(l)}_{kj}\}$.
\State \textbf{Output}: $\hat{\Omega}^{(l)}=[\hat{\bs{\theta}}^{(l)}_1,\ldots,\hat{\bs{\theta}}^{(l)}_p]$
\end{algorithmic}
\end{algorithm}
\end{document}